\documentclass[aps,prb,twocolumn,floats,fleqn,eqsecnum]{revtex4}
\usepackage{amsmath}
\usepackage{graphicx}

\newcommand{\Tr}{\operatorname{Tr}}

\begin{document}

\title{Equation of Motion Method for Composite Field Operators}
\author{Ferdinando Mancini}
\email[E-mail: ]{mancini@sa.infn.it}
\author{Adolfo Avella}
\email[E-mail: ]{avella@sa.infn.it}
\affiliation{Dipartimento di Fisica ``E.R. Caianiello'' - Unit\`a di Ricerca INFM di Salerno\\
Universit\`a degli Studi di Salerno, I-84081 Baronissi (SA),Italy}
\date{September 27, 2003}

\begin{abstract}
The Green's function formalism in Condensed Matter Physics is
reviewed within the equation of motion approach. Composite
operators and their Green's functions naturally appear as building
blocks of generalized perturbative approaches and require fully
self-consistent treatments in order to be properly handled. It is
shown how to unambiguously set the representation of the Hilbert
space by fixing both the unknown parameters, which appear in the
linearized equations of motion and in the spectral weights of
non-canonical operators, and the zero-frequency components of
Green's functions in a way that algebra and symmetries are
preserved. To illustrate this procedure some examples are given:
the complete solution of the two-site Hubbard model, the
evaluation of spin and charge correlators for a narrow-band Bloch
system, the complete solution of the three-site Heisenberg model,
and a study of the spin dynamics in the Double-Exchange model.
\end{abstract}


\maketitle

\section{Introduction}

\label{Intro}

The physical system analyzed in this paper is an aggregate of
interacting \emph{Wannier}-electrons \emph{living} on a lattice
spanned by the vectors $ \mathbf{i}$. For the sake of simplicity,
we restrict our study to single-band electron models; the
generalization to multi-band models is straightforward. The system
is enclosed in a finite, but macroscopically large, volume $V$,
containing $M$ sites of the lattice, and is supposed to be in
thermodynamic equilibrium at a temperature $T$. In a
second-quantization scheme the dynamics of this system is ruled by
a certain Hamiltonian $\hat{H}=\hat{H}\left[\varphi(i)\right]$
describing, in complete generality, the free propagation of the
electrons and all the interactions among them and with external
fields (e.g., electromagnetic fields, pressure and temperature
gradients,...). $\varphi(i)$ denotes an Heisenberg electronic
field [$i=\left(\mathbf{i},t\right)$] in spinorial notation
satisfying canonical anticommutation relations. Any physical
property of this system can be connected to the expectation value
of a specific operator.

The expectation value $\langle \hat{A}\rangle$ of any operator
$\hat{A}= \hat{A}\left[\varphi(i)\right]$ can be computed, for the
grand-canonical ensemble, by taking the normalized trace of the
operator weighted with the quantum-mechanical statistical factor
$\mathrm{e}^{-\beta\left(\hat{H}-\mu \hat{N}\right)}$.
$\hat{N}=\sum_{\mathbf{i}\sigma}\varphi_{\sigma}^{\dag}(i)
\varphi_{\sigma}(i)$ is the total number operator, $\beta$ is the
inverse temperature and $\mu$ is the chemical potential, which is
fixed in order to get the desired average number of particles
$N=\langle \hat{N}\rangle$. The chemical potential will be a
function of $N$ and $T$, as well as other parameters present in
the Hamiltonian. Although the trace can be taken over any basis
(i.e., over any complete set of states in the Hilbert space of the
system), the most convenient one, the \textbf{eigenbasis}, is
constituted by the simultaneous eigenstates of $\hat{H}$ and
$\hat{N}$. If such a basis is known, then all the properties of
the system can be exactly calculated: this procedure is known as
exact diagonalization (\emph{ED}). It is worth reminding that the
Hilbert space of a fermionic system contains only those states
compatible with the Pauli principle (i.e., states with occupation
numbers per site and spin\cite{Pauli1} equal to either 0 or 1).

Generally, \emph{ED} can be effectively applied only to systems
that are non-interacting or interacting, but very small. In
particular, if the system is non-interacting the eigenbasis
coincides with the canonical basis of the Fock space of the system
(i.e., the set of states constructed by locating the electrons,
one at a time, on the lattice sites in accordance with the Pauli
principle). For small systems, it is always possible to exactly
diagonalize the Hamiltonian according to the reasonable small
number of available states, but when large interacting systems are
considered the number of states can be enormous and \emph{ED} is
practically not applicable. This consideration gave birth to
numerous numerical techniques: Lanczos, quantum Monte Carlo,...,
which can be considered as attempts to construct an approximate
version of \emph{ED} that could be applied to very large systems.
However, these numerical techniques have some very severe
limitations coming from the unavoidable small number of sites they
can treat (the computational time increases exponentially with the
number of available states): they cannot give a reliable
description of systems with long range interactions; phases
presenting long range order of any kind are absolutely
unaccessible; the very low resolution in frequency and momentum
prevents the applications to systems with relevant low-energy
features (e.g., systems that present Kondo-like effects) or with
strong spatial-dependence or anisotropy in their physical
properties (e.g., systems that have a Fermi surface ill-defined,
nodal or with high angular-momentum symmetry). Moreover, the
information we get by means of these techniques for a system of a
certain size difficultly can be used for a system of bigger size
and, even worse, does not give any clear idea of what can happen
in the corresponding bulk system.

According to this, we have to find an alternative exact analytical
technique that can generate, for large interacting systems,
approximate treatments not suffering from the very severe
limitations noticed in the numerical methods. In principle, this
technique will obviously give the same exact results of \emph{ED}.
Coming back to our original problem, the evaluation of the
expectation value $\langle \hat{A}\rangle$, it is possible to use
the equation of motion
\begin{equation}  \label{Eq2.6}
\mathrm{i}\frac{\partial}{\partial t}
\varphi(i)=\left[\varphi(i),\hat{H} \right]
\end{equation}
in order to derive one or more equations for this quantity or,
better, for the corresponding Green's function (see next section).
Actually, the equation of motion (\ref{Eq2.6}) naturally generates
\textbf{higher-order operators} (i.e., operators constituted by
more and more elementary fields, some of them centered on farther
and farther sites from $\mathbf{i}$). The process can be iterated
by time-differentiating the newly generated operators and a chain
of equations of motion can be constructed. The obtained system of
equations of motion closes on a complete set of
\textbf{eigenoperators} of the Hamiltonian
\begin{equation}  \label{Eq2.7}
\mathrm{i}\frac{\partial}{\partial t}
\psi\left(\mathbf{i},t\right)=\left[
\psi\left(\mathbf{i},t\right),\hat{H}\right] = \sum_{\mathbf{j}}
\varepsilon\left(\mathbf{i},\mathbf{j}\right)
\psi\left(\mathbf{j},t\right)
\end{equation}
where $\psi(i)$ is a $n$-component spinorial field and
$\varepsilon\left( \mathbf{i},\mathbf{j}\right)$, usually called
the \textbf{energy matrix}, is a square matrix of rank $n$. This
approach is known as \textbf{the equations of motion method}
(\emph{EM}) and can be applied, obviously giving the same exact
results, in all cases where we can also apply \emph{ED}: if the
system is non-interacting the original electronic operators
$\varphi(i)$ are the eigenoperators of the Hamiltonian; for small
systems the number of equations of motion to be solved
simultaneously, in order to find the complete set of
eigenoperators, is reasonably small and makes the application
feasible. When large interacting systems are considered the number
of eigenoperators rapidly increases (diverges in the thermodynamic
limit) and \emph{EM} cannot be effectively applied just as
\emph{ED} could not be. However, the main difference between the
two procedures is that \emph{EM} can be still used in some
approximation not subject to the severe limitations noticed in the
numerical techniques derived from \emph{ED}.

Any approximation derived from \emph{EM} is based on some of the
peculiar properties of eigenoperators (some of them are reported
below). These properties are obviously not enjoined by eigenstates
that have to be considered always as a whole (symmetry
considerations can only reduce a brute force diagonalization to a
more refined block diagonalization which is in any way unfeasible
in the thermodynamic limit). The iterated process of
time-differentiation generates more and more \emph{delocalized}
eigenoperators in direct space (i.e., eigenoperators containing
original fields siting on more and more distant sites), which are
less and less relevant as they have eigenenergies rapidly
decreasing with the \textbf{spatial size of the eigenoperator}
(i.e., with the maximum distance among the sites where the
constituting original fields are sited). Although the total number
of eigenoperators is equivalent to the number of possible
transitions among all the eigenstates and, therefore, goes as this
latter number squared (i.e., if the number of eigenstates is $n$,
then the number of eigenoperators is $ \frac{n(n+1)}2$), to study
a specific physical property we need only to analyze the dynamics
of the few eigenoperators relevant to it. Furthermore, the
eigenoperators can be easily generalized to any size of the system
and the dynamics of all sites can be studied at once; this is
impossible for the eigenstates. According to this, for very small
clusters too, where the application of \emph{ED} requires
undoubtedly less effort than that of \emph{EM}, \emph{EM} solution
is preferable as it has the fundamental property to be
\textbf{scalable} (i.e., it gives a lot of information about both
\emph{EM} solution of bigger clusters and the approximate
\emph{EM} solution of the corresponding bulk system).

The line of thinking described so far follows the developments of
the condensed matter physics in the last decades. Both \emph{ED}
and \emph{EM} try to \emph{diagonalize} the Hamiltonian under
study, but in two different spaces. The former searches for the
eigenbasis within the Hilbert space of the system, the latter
seeks an \textbf{operatorial basis} within the field space
generated by the application of the Hamiltonian to the original
field and to its \emph{bosonic} aggregations (i.e., to fields
constructed by an even number of original fields). While the
states of a system drastically change with its size (i.e., the
corresponding Hilbert spaces do not overlap), the operators just
increase in number and complexity (i.e., the new field space
include the old one). Moreover, the relevance of an eigenoperator,
which is \emph{measured} by the magnitude of the scale of energy
it describes, usually survives any change in size of the system.
We can also define as \textbf{minimal cluster} the smallest one
allowing all the terms of the Hamiltonian to act properly. Only
eigenoperators obtained for systems realized on clusters at least
equal to the minimal one can be trusted and used to describe
properties of the corresponding bulk systems.

In order to construct any approximation scheme in the framework of
\emph{EM}, a convenient generalization of the concept of
correlation function is provided by that of Green's function
\cite{Bogoliubov:59} (\emph{GF}). The latter has some advantages
in the construction and solution of the equations that determine
it. Moreover, the \emph{GF} contain most of, practically all, the
relevant information on the properties of the system: expectation
values of observables, excitation spectrum, response to external
perturbations, and so on. Different types of \emph{GF} can be
constructed; we will consider real-time thermodynamic \emph{GF}
where the thermal averaging process of the Heisenberg operators is
performed over the grand-canonical ensemble.

The traditional approximation schemes, often based on perturbative
calculations, use as building blocks the non-interacting
\emph{GF}. The mean-field formulation, which corresponds to the
linearization of the equation of motion (\ref{Eq2.6}) [i.e.,
$\mathrm{i}\frac{\partial}{\partial t}
\varphi\left(\mathbf{i},t\right)= \sum_{\mathbf{j}}
\varepsilon\left( \mathbf{i},\mathbf{j}\right)
\varphi\left(\mathbf{j},t\right)$, where $
\varepsilon\left(\mathbf{i},\mathbf{j}\right)$ is now a scalar
function], also belongs to this category. An intense study has
been performed along this line and many techniques have been set
up: perturbation expansions on the basis of Feynman diagrams,
Dyson equation, Wick's theorem, and so on. It is worth noting that
in order to describe phases with different symmetries, these
schemes need to become self-consistent.

All these techniques rely on the hypothesis that the interactions
among the electrons are weak and can be treated in the framework
of some perturbation scheme. However, as many and many theoretical
and experimental studies have shown with more and more convincing
evidence, all these methods are not adequate to treat strongly
correlated electron systems (\emph{SCES}) and different approaches
must be considered. In these systems, the fundamental concept of
the electron as a particle with some well-defined properties
completely breaks down. The presence of the correlations modifies
the properties of the electrons and, at the macroscopic level, new
particles are observed, with peculiar properties entirely
determined by the dynamics and the \textbf{boundary conditions}
(i.e., all the elements characterizing the physical situation we
wish to study). These new objects appear as the final result of
the modifications imposed on the electrons by the interactions and
contain, by the very beginning, a relevant part of the effects of
correlation.

As simple, but significative, example, let us consider an atomic
model with a local interaction $U$ between the electrons (i.e.,
$\hat{H} =U\varphi_{\uparrow}^{\dag}\varphi_{\uparrow}
\varphi_{\downarrow}^{\dag}\varphi_{\downarrow}$). This model is
exactly solvable in terms of the Hubbard operators
\begin{equation}  \label{Eq2.8}
\begin{split}
&\xi=\left[1-\varphi^{\dag}\varphi\right]\varphi \\
&\eta=\left[\varphi^{\dag}\varphi\right]\varphi
\end{split}
\end{equation}
Due to the presence of the local interaction $U$, the original
electrons $ \varphi(i)$ are no more observables and new stable
elementary excitations, described by the field operators $\xi(i)$
and $\eta(i)$, appear.

On the basis of this evidence, one can be induced to move the
attention from the original fields to the new fields generated by
the interactions. The operators describing these excitations can
be written in terms of the original ones and are known as
\textbf{composite operators}. Several approaches have been
formulated where composite fields are used as operatorial basis
for developing approximation
schemes\cite{Mori,Rowe:68,Roth:69,Nolting:72,Tserkovnikov:81,Nolting:89,Plakida:89,Fedro:92,Fulde:95,Mancini-IJMPB,Avella:98,Matsumoto}.
All these approaches are very promising: some amount of the
interaction is already present in the chosen basis and this
permits to overcome the problem of finding an appropriate
expansion parameter. However, a price must be paid. In general,
the composite fields are neither Fermi nor Bose operators, since
they do not satisfy canonical (anti)commutation relations, and
their properties, because of the inherent definition, must be
self-consistently determined. They can only be recognized as
fermionic or bosonic according to the number of constituting
original particles.

New techniques of calculus have to be used in order to deal with
composite fields. In developing approximation schemes where the
building blocks are now the propagators of composite operators,
one cannot use the standard version of the consolidated schemes;
diagrammatic expansions, Wick's theorem and many other
prescriptions are no more valid for composite operators. There
have been attempts\cite{Vaks:62,Zaitsev:76,Izyumov:90} to extend
these schemes, but although very good results have been obtained
for spin operators\cite{Vaks:62}, the complexity (and often the
ambiguity) of the analytical calculations required by the Hubbard
operators (the simpler among the fermionic composite operators)
does not allow, at least at the present, an effective application
of such techniques to real problems. The formulation of the
\emph{GF} method must be revisited. As it will be shown below,
three serious problems arise when we wish to study the propagators
of composite fields:

\begin{enumerate}
\item the appearance of some unknown parameters as correlation
functions of field operators not belonging to the chosen
operatorial basis;\label{1prob}

\item the appearance of some zero-frequency constants (\emph{ZFC})
as a consequence of the existence of zero-frequency
modes;\label{2prob}

\item the necessity of fixing the representation where the
\emph{GF} are formulated.\label{3prob}
\end{enumerate}

In most of the approaches found in the literature the solution to
the previous problems is the following.

\ref{1prob}. In order to determine the unknown parameters several
methods (arbitrary ansatz, decoupling schemes, use of the equation
of motion,...) have been considered in the context of different
approaches (Hubbard I approximation, Roth's method, projection
method, spectral density approach,...). All these methods suffer
from the severe limitation of not being fully self-consistent. On
the other hand, any approach based on the correct use of composite
operators is, by construction, a fully self-consistent approach.
As shown in Ref.~\onlinecite{Avella:98}, in the context of the
Hubbard model, all these procedures lead to a series of unpleasant
results: several sum rules and the particle-hole symmetry are
violated, there is no presence of a Mott transition, all local
quantities strongly disagree with the results of the numerical
simulation.

\ref{2prob}. Any symmetry enjoined by the Hamiltonian induces a
degeneracy among the eigenstates of the system. The equivalence of
two or more eigenenergies implies the presence of zero-energy
modes. In the case of bosonic Green's functions these modes give
rise to some unknown quantities that we will call \emph{ZFC}. The
\emph{ZFC} are really relevant quantities as they are connected to
fundamental physical properties such as the compressibility and
the specific heat: they can be considered as a measure of the
fluctuations, quantum and/or thermal ones, present in the thermal
averages of the generators of the symmetry group, which are
usually bosonic. The \emph{ZFC} are usually fixed by requiring the
ergodicity of the dynamics of the relative operators with respect
to the Hamiltonian under study. This is clearly a very strong
assumption. As it will be shown in the third section of this
paper, there are non-trivial examples of exactly-solvable systems
where the \emph{ZFC} do not assume their ergodic value: if we
would have forced the \emph{ZFC} to assume it, this would have
implied a zero compressibility, specific heat,... Furthermore,
although the response functions do not explicitly depend on them,
there is an implicit dependence due to the inherent
self-consistency of the entire scheme. According to this, in
general, these quantities must be calculated case by case.

\ref{3prob}. The knowledge of the Hamiltonian and of the
operatorial algebra is not sufficient to completely specify the
\emph{GF}. The \emph{GF} refer to a specific
\textbf{representation} (i.e., to a specific choice of the Hilbert
space) and this information must be supplied to the equations of
motion that alone are not sufficient to completely determine the
\emph{GF}. As well known, the same system can exist in different
phases according to the external conditions; the existence of
infinite inequivalent representations \cite{Umezawa:93} where the
equations of motions can be realized, allows us to pick up, among
the many possible choices, the right Hilbert space appropriate to
the physical situation under study. The construction of the
Hilbert space where the \emph{GF} are realized is not an easy task
and is usually ignored. The use of composite operators leads to an
enlargement of the Hilbert space by the inclusion of some
unphysical states. As a consequence of this, it is difficult to
satisfy a priori all the sum rules and, in general, the symmetry
properties enjoined by the system under study. In addition, since
the representation where the operators are realized has to be
dynamically determined, the method clearly requires a process of
self-consistency.

In the Composite Operator Method\cite{Mancini-IJMPB,Avella:98}
(\emph{COM}), as illustrated in the next Section, the three
problems are not considered separately but they are all connected
in one self-consistent scheme. The main idea is that fixing the
values of the unknown parameters and of the \emph{ZFC} implies to
put some constraints on the representation where the \emph{GF} are
realized. As the determination of this representation is not
arbitrary, it is clear that there is no freedom in fixing these
quantities. They must assume values compatible with the dynamics
and with the right representation. Which is the right
representation? This is a very hard question to answer. From the
algebra it is possible to derive several relations among the
operators (e.g., $\varphi_{\sigma}(i)\varphi_{\sigma}(i)=0$): we
will call them Algebra Constraint relations (\emph{AC}). This set
of relations, valid at microscopic level, must be satisfied also
at macroscopic level (i.e., when the expectations values are
considered; e.g., $\left\langle\varphi_{\sigma}(i)\varphi_{
\sigma}(i)\right\rangle=0$). We also note that in general the
Hamiltonian has some symmetry properties (e.g., rotational
invariance in coordinate and spin space, phase invariance, gauge
invariance,...). These symmetries generate a set of relations
among the $n$-point Green's functions: the Ward-Takahashi
relations\cite{WT} (\emph{WT}). It is worth noting that many
approximations present in the literature do not fulfill these
consistency requirements and, consequently, obtain \textsl{wrong}
results. Now, certainly the right representation must be the one
where all relations among the operators satisfy the conservation
laws present in the theory when expectation values are taken
(i.e., where all the \emph{AC} and \emph{WT} are preserved). Then,
we impose these conditions and obtain a set of self-consistent
equations that will fix the unknown correlators, the \emph{ZFC}
and the right representation at the same time, avoiding the
problem of \emph{uncontrolled} and \emph{uncontrollable}
decouplings, which affects many different approximation schemes
and has been here definitely solved. This is the main ingredient
of the \emph{COM}, together with the recipes\cite{recipe} that we
have developed in the last years in order to choose the
appropriate operatorial basis of composite operators according to
the specific system under analysis. As regards this last issue, we
wish to drive the attention on the procedure we propose as it can
be considered a systematic attempt to seek and build up (exact as
much as it is possible) operatorial basis for interacting systems.
This is a new frontier in condensed matter theory (quantum Hall
effect, heavy-fermion systems, quantum critical points, competing
unconventional ordering phenomena, breakdown of Fermi liquid
picture, connections among spin, charge, orbital and lattice
degrees of freedom, ...) and our procedure should be regarded as
an attempt to revisit the established picture for strongly
correlated systems.

The second section is devoted to revisit the \emph{GF} formalism
in presence of composite fields and to establish the \emph{COM} as
a general procedure to compute \emph{GF} of highly correlated
systems. In the third section of the paper we will illustrate the
formalism by considering some specific examples: the two-site
Hubbard model, the three-site Heisenberg model, a narrow-band
Bloch system in presence of an external magnetic field and the
double-exchange model. For the two-site Hubbard model we compute
the fermionic \emph{GF} independently from the bosonic one by
means of the \emph{AC}. The latter also allow us to fix the
\emph{ZFC} of the bosonic \emph{GF}, which result in not being
ergodic, and to get straightforwardly the right representation.
The solution of the tree-site Heisenberg model shows the
impossibility to get any spontaneously ordered state at finite
temperature in a finite system as a consequence of internal
consistency in the proposed formulation. Moreover, it is really
relevant the existing relation between the number of \emph{ZFC}
appearing in the \emph{GF} and the presence of the magnetic field.
In the case of a narrow-band Bloch system in presence of an
external magnetic field we will see that the \emph{ZFC} relative
to the total number operator, which is an integral of motion, has
a non-ergodic value, even if we have an ergodic charge dynamics.
The double-exchange model finally gives us the possibility to show
one way to apply the proposed formulation to large interacting
systems. In this case, we also show how to recognize the
manifestation of the Mermin-Wagner theorem\cite{Mermin:66} within
this formulation. For the exactly-solvable models used in the
examples (i.e., the Hubbard and the Heisenberg models), we have
checked, although it could not be otherwise, that the proposed
formulation reproduces the exact results coming from \emph{ED}.
Actually, the \emph{COM} has been developed just for systems large
and interacting; the applications to systems that are small or
non-interacting have only to be interpreted as mere demonstrations
of all features of the method and of its power and correctness.
Finally, in the Appendices, we give: a generalized perturbative
approach for strongly interacting systems; part of the derivation
of the general formulation; the derivation of the zero-temperature
formulation; the \emph{GF} expressions, dispersion relations and
sum rules where the presence of the \emph{ZFC} is explicitly taken
into account.

\section{General Formalism}

\label{Forma}

This section is devoted to revisit the \emph{GF} formalism in
presence of composite fields and to establish the \emph{COM} as a
general procedure to compute \emph{GF} of complex interacting
systems. Owing to the difficulties in dealing with composite
operators, reported in detail in the previous section, the study
is performed completely within \emph{EM}. We start by considering
a set of composite fields, chosen according to a well-defined
recipe\cite{recipe}. The fields can be of fermionic or bosonic
nature, according to the physical properties we wish to
study\cite{basis}. In the case of fermionic operators it is
intended that we use the spinorial representation. The set
$\psi(i)$ satisfies a linear system of equations of motion [see
Eq.~(\ref{Eq2.7})]. If the fields $\psi(i)$ are eigenoperators of
the total Hamiltonian, the equations of motion are exact. Several
examples will be given in Sec.~III. If the fields $\psi(i)$ are
not eigenoperators of the Hamiltonian, the equations of motion are
approximated and all the formalism is developed with the aim of
computing and using the propagators of these fields as a basis to
set up a perturbative scheme of calculations. In Appendix~A, we
give a sketch of a generalized perturbative approach based on a
Dyson equation (Eq.~\ref{Dyson}) designed for formulations using
composite fields. Then, the total weight of the self-energy
corrections is bounded by the \textbf{weight} of the residual
source operator $\delta\!J(i)$ [see Eq.~(\ref{EMexp})]. According
to this, it can be made smaller and smaller by increasing the
components of the basis $\psi(i)$ (e.g., by including higher-order
composite operators appearing in $\delta\!J(i)$). The result of
such procedure will be the inclusion in the energy matrix of part
of the self-energy as an expansion in terms of coupling constants
multiplied by the weights of the newly included basis operators.
In general, the enlargement of the basis leads to a new
self-energy with a smaller total weight. However, it is necessary
pointing out that this process can be quite cumbersome and the
inclusion of fully momentum and frequency dependent self-energy
corrections can be necessary to effectively take into account
low-energy and virtual processes. According to this, one can
choose a reasonable number of components for the basic set and
then use another approximation method to evaluate the residual
dynamical corrections (e.g., specially adapted versions of the
non-crossing approximation or of the FLEX).

By considering two-time thermodynamic
\emph{GF}\cite{Bogoliubov:59,Zubarev:60,Zubarev:74}, let us define
the causal function
\begin{multline}  \label{Eq1.2}
G^{(\eta)}_C(i,j) = \theta\!\left(t_i-t_j\right) \left\langle
\psi(i)\,\psi^\dag(j)\right\rangle \\
-\eta\,\theta(t_j-t_i) \left\langle
\psi^\dag(j)\,\psi(i)\right\rangle
\end{multline}
the retarded and advanced functions
\begin{equation}  \label{Eq1.3}
G^{(\eta)}_{R,A}(i,j) = \pm
\theta\!\left[\pm\left(t_i-t_j\right)\right] \left\langle
\left[\psi(i),\,\psi^\dag(j)\right]_{\eta}\right\rangle
\end{equation}
and the correlation function
\begin{equation}  \label{Eq1.4}
C(i,j)=\left\langle \psi(i)\,\psi^\dag(j)\right\rangle
\end{equation}
Here $\eta=\pm1$; usually, it is convenient to take $\eta=1$
($\eta=-1$) for a \emph{fermionic} (\emph{bosonic}) set $\psi(i)$
(i.e., for a composite field constituted of an odd (even) number
of original fields $\varphi(i)$) in order to exploit the canonical
anticommutation relations satisfied by $ \varphi(i)$; but, in
principle, both choices are possible. Accordingly, we define
\begin{equation}  \label{Eq1.5}
\left[A,\,B\right]_{\eta}=
\begin{cases}
\left\{A,\,B\right\}=A\,B+B\,A & \text{for $\eta=1$} \\
\left[A,\,B\right]=A\,B-B\,A & \text{for $\eta=-1$}
\end{cases}
\end{equation}
$<\cdots>$ denotes the quantum-statistical average over the grand
canonical ensemble. From the equation~(\ref{Eq2.7}) for the set
$\psi(i)$, the Fourier transforms of these functions satisfy the
following equations (we consider a translational invariant system)
\begin{subequations}
\label{Eq1.6}
\begin{align}
&\left[\omega-\varepsilon(\mathbf{k})\right]
G^{(\eta)}_{C,R,A}(\mathbf{k}
,\omega)=I^{(\eta )}(\mathbf{k}) \\
&\left[\omega-\varepsilon(\mathbf{k})\right]
C(\mathbf{k},\omega)=0
\end{align}
where
\end{subequations}
\begin{equation}  \label{Eq1.7}
I^{(\eta)}(\mathbf{k}) = \mathcal{F} \left\langle
\left[\psi(\mathbf{i} ,t),\,
\psi^\dag(\mathbf{j},t)\right]_{\eta}\right\rangle
\end{equation}
is known as the \textbf{normalization matrix}. $\mathcal{F}$
indicates the Fourier transform. The most general solution of
equations~(\ref{Eq1.6}) is
\begin{subequations}
\label{Eq1.8}
\begin{multline}
G^{(\eta)}_{C,R,A}\left(\mathbf{k},\omega\right) = \sum_{l=1}^n
\left\{
\mathcal{P}\left[\frac{\sigma^{(\eta,l)}(\mathbf{k})}{\omega
-\omega_l(
\mathbf{k})}\right]\right. \\
\left. - \mathrm{i}\,\pi\,\delta\!\left[\omega
-\omega_l(\mathbf{k})\right]
\,g^{(\eta,l)}_{C,R,A}(\mathbf{k})\right\}
\end{multline}
\begin{equation}
C\left(\mathbf{k},\omega\right) = \sum_{l=1}^n\delta\!\left[\omega
-\omega_l( \mathbf{k})\right]\,c^{(l)}(\mathbf{k})
\end{equation}
$g^{(\eta,l)}_{C,R,A}(\mathbf{k})$ and $c^{(l)}(\mathbf{k})$ are
not fixed by the equations of motion and have to be determined by
imposing the appropriate \emph{boundary conditions}.
$\omega_l(\mathbf{k})$ are the eigenvalues of the matrix
$\varepsilon(\mathbf{k})$. $\sigma^{(\eta,l)}( \mathbf{k})$ are
the spectral density functions and can be expressed in terms of
the matrices $\varepsilon(\mathbf{k})$ and
$I^{(\eta)}(\mathbf{k})$ as
\end{subequations}
\begin{equation}  \label{Eq1.9}
\sigma_{\alpha \beta}^{(\eta,l)}(\mathbf{k})=\Omega _{\alpha
l}(\mathbf{k} )\sum_\delta\Omega _{l\delta
}^{-1}(\mathbf{k})\,I_{\delta \beta}^{(\eta )}( \mathbf{k})
\end{equation}
where $\Omega(\mathbf{k})$ is the $n\times n$ matrix whose columns
are the eigenvectors of the matrix $\varepsilon(\mathbf{k})$. The
summations run over the number of eigenvalues of
$\varepsilon(\mathbf{k})$ and $\mathcal{P}$ represents the
principal value.

By recalling the \emph{boundary conditions} $G^{(\eta)}_R(t<0)=0$
and $ G^{(\eta)}_A(t>0)=0$ it is immediate to see that
\begin{equation}  \label{Eq1.10}
g^{(\eta,l)}_R(\mathbf{k})=-g^{(\eta,l)}_A(\mathbf{k})=\sigma
^{(\eta,l)}( \mathbf{k})
\end{equation}
Then, the retarded and advanced \emph{GF} are completely
determined in terms of the matrices $\varepsilon(\mathbf{k})$ and
$I^{(\eta)}(\mathbf{k})$.

The determination of $g^{(\eta,l)}_C(\mathbf{k})$ and
$c^{(l)}(\mathbf{k})$ require some more work. On the basis of the
calculations reported in App.~B, it is straightforward to obtain
the following results
\begin{subequations}
\label{Eq1.16}
\begin{align}
& \sum_{l\in \mathcal{A}(\mathbf{k})}g_{C}^{(\eta
,l)}(\mathbf{k})=(1-\eta
)\Gamma (\mathbf{k})  \label{Eq1.16a} \\
& \sum_{l\in \mathcal{A}(\mathbf{k})}\sigma
^{(1,l)}(\mathbf{k})=(1+\eta
)\Gamma (\mathbf{k})  \label{Eq1.16b} \\
& c^{(l)}(\mathbf{k})=\frac{2\pi}{1+\eta \,e^{-\beta \,\omega
_{l}(\mathbf{k})}}\sigma ^{(\eta
,l)}(\mathbf{k})\,\,\,\,\,\,\forall l\in \mathcal{B}(
\mathbf{k})  \label{Eq1.16d} \\
& g_{C}^{(\eta ,l)}(\mathbf{k})=\frac{1-\eta \,e^{-\beta \,\omega
_{l}( \mathbf{k})}}{1+\eta \,e^{-\beta \,\omega
_{l}(\mathbf{k})}}\sigma ^{(\eta
,l)}(\mathbf{k})\,\,\,\,\,\,\forall l\in \mathcal{B}(\mathbf{k})
\label{Eq1.16e}
\end{align}
where $\mathcal{A}(\mathbf{k})$ and $\mathcal{B}(\mathbf{k})$ are
explicitly defined in Eq.~(\ref{momreg}) and $\Gamma
(\mathbf{k})$, \textbf{the zero-frequency function}, is defined as
\end{subequations}
\begin{equation}
\Gamma (\mathbf{k})=\frac{1}{2\pi}\sum_{l\in
\mathcal{A}(\mathbf{k} )}c^{(l)}(\mathbf{k}) \label{Eq1.16def}
\end{equation}

We see that Eq.~(\ref{Eq1.16b}) requires that
\begin{equation}
\sum_{l\in \mathcal{A}(\mathbf{k})}\sigma ^{(-1,l)}(\mathbf{k})=0
\label{Eq1.16c}
\end{equation}
This condition comes from the requirement that the correlation
function in direct space should not diverge: a solution with
$\sum_{l\in \mathcal{A}( \mathbf{k})}\sigma
^{(-1,l)}(\mathbf{k})\neq 0$ implies a divergence of the Fourier
coefficients $c^{(l)}(\mathbf{k})$ for any finite temperature.
This is admissible only if the divergence is integrable and the
corresponding direct space correlation function remains finite. A
finite value of $\sum_{l\in \mathcal{A}(\mathbf{k})}\sigma
^{(-1,l)}(\mathbf{k})$ is generally related to the presence of
long-range order (i.e., symmetry breaking) and the previous
statement is nothing but the Mermin-Wagner
theorem\cite{Mermin:66}. A detailed analysis of this point will be
illustrated in Section III by investigating the Heisenberg and
Double Exchange Models.

By putting Eqs.~(\ref{Eq1.10}) and (\ref{Eq1.16}) into
Eqs.~(\ref{Eq1.8}) we get the following general expressions for
the \emph{GF}
\begin{subequations}
\label{Eq1.18}
\begin{equation}
G^{(\eta )}_{R,A}(\mathbf{k},\omega) = \sum_{l\in\aleph}
\frac{\sigma ^{(\eta,l)}(\mathbf{k})}{\omega - \omega
_l(\mathbf{k})\pm \mathrm{i} \,\delta}
\end{equation}
\begin{multline}
G^{(\eta)}_C(\mathbf{k},\omega ) =
\Gamma(\mathbf{k})\left(\frac1{\omega+
\mathrm{i}\,\delta}+\frac{\eta}{\omega-\mathrm{i}\,\delta}\right) \\
+\sum_{l\in\mathcal{B}(\mathbf{k})} \frac{\sigma
^{(\eta,l)}(\mathbf{k})} { 1+\eta\,e^{-\beta
\omega}}\left[\frac{1}{\omega -\omega_l(\mathbf{k})+
\mathrm{i}\,\delta} + \frac{\eta\,e^{-\beta\,\omega}}{\omega
-\omega_l( \mathbf{k})-\mathrm{i}\,\delta}\right]
\end{multline}
\begin{multline}  \label{Eq1.18c}
C(\mathbf{k},\omega )= 2\pi\,\delta(\omega)\,\Gamma(\mathbf{k}) \\
+ 2 \pi \sum_{l\in\mathcal{B}(\mathbf{k})} \delta\!\left[\omega
-\omega_l( \mathbf{k})\right]\frac{\sigma ^{(\eta,l)}(\mathbf{k})}
{1+\eta\,e^{-\beta\, \omega_l(\mathbf{k})}}
\end{multline}
As shown in Appendix~C, Eqs.~(\ref{Eq1.18}) hold also in the limit
of zero temperature (i.e., in the limit $\beta\rightarrow\infty$).
From these expressions it is possible to get dispersion relations
and sum rules that take explicitly into account the presence of
the zero-frequency function (see Appendix~D).

We see that the general structure of the \emph{GF} is remarkably
different according to the statistics. For \emph{fermionic}
composite fields (i.e., when it is natural to choose $\eta=1$) the
zero-frequency function $\Gamma( \mathbf{k})$ contributes to the
spectral function, it is directly related to the spectral density
functions by means of equation (\ref {Eq1.16b}) and its
calculation does not require more information. For \emph{bosonic}
composite fields (i.e., when it is natural to choose $\eta=-1$)
the zero-frequency function does not contribute to the spectral
function, but to the imaginary part of the causal \emph{GF}. The
causal and retarded (advanced) \emph{GF} contain different
information and the right procedure of calculation is controlled
by the statistics. In particular, in the case of bosonic fields
(i.e., for $\eta=-1$) one must start from the causal function and
then use
\end{subequations}
\begin{equation}  \label{Eq1.34}
\begin{split}
&\Re\left[G^{(-1)}_{R,A}(\mathbf{k},\omega
)\right]=\Re\left[G^{(-1)}_C(
\mathbf{k},\omega )\right] \\
&\Im\left[G^{(-1)}_{R,A}(\mathbf{k},\omega )\right]=\pm \tanh
\frac{
\beta\,\omega}{2}\,\Im\left[G^{(-1)}_C(\mathbf{k},\omega )\right] \\
&C(\mathbf{k},\omega )=-\left[1+\tanh
\frac{\beta\,\omega}{2}\right] \Im
\left[G^{(-1)}_C(\mathbf{k},\omega )\right]
\end{split}
\end{equation}
On the contrary, for fermionic fields (i.e., for $\eta=1$) the
right procedure for computing the correlation function requires
first the calculation of the retarded (advanced) function and then
the use of relations identical to those of Eqs.~\ref{Eq1.34}, but
with the subscript $R,A$ and $C$ inverted and the minus sign in
the last equation changed to $\mp$.

Moreover, it is worth noting that $\Gamma(\mathbf{k})$ is
undetermined within the \emph{bosonic} sector (i.e., for
$\eta=-1$) and should be computed in the \emph{fermionic} sector
(i.e., for $\eta=1$) by means of equation (\ref{Eq1.16b}) or
equivalently by means of the following relation
\begin{equation}  \label{Eq1.34b}
\Gamma(\mathbf{k})=\frac12 \lim_{\omega\rightarrow0}
\omega\,G^{(1)}_C( \mathbf{k},\omega )
\end{equation}
However, the calculation of $\sigma^{(1,l)}(\mathbf{k})$ requires
the calculation of $I^{(1)}(\mathbf{k})$ that, for \emph{bosonic}
fields, generates unknown momentum dependent correlation functions
whose determination can be very cumbersome as requires, at least
in principle, the self-consistent solution of the integral
equations connecting them to the corresponding Green's functions.
In practice, also for simple, but anyway composite, \emph{bosonic}
fields the $\Gamma(\mathbf{k})$ remains undetermined and other
methods rather than equation (\ref{Eq1.16b}) should be used.
Similar methods, like the use of the relaxation
function\cite{Kubo:57}, would lead to the same problem.

The zero-frequency function $\Gamma(\mathbf{k})$ is known in the
literature
\cite{Kubo:57,Callen:67,Suzuki:71,Huber:77,Aksenov:78,Aksenov:78a,Aksenov:87}
as an indicator of the ergodic nature of the dynamics of the
operator $ \psi(i)$ with respect to the Hamiltonian $\hat{H}$. We
recall that a quantity $A$ has an ergodic dynamics if and only if
\begin{equation}  \label{Eq1.35}
\lim_{t \rightarrow \infty}\left\langle A(t)\,A\right\rangle
=\left\langle A\right\rangle ^2
\end{equation}
that is, if and only if its auto-correlation attenuates in the
time. We have not to forget that the condition (\ref{Eq1.35}) is
the same as the standard ergodic requirement (i.e., equivalence of
averages taken in time and over the phase space) only for
statistical averages computed in the microcanonical
ensemble\cite{Kubo:57}; in other ensembles it holds only in the
thermodynamic limit. By recalling the general
expression~(\ref{Eq1.18c}) for the correlation function, the
condition of ergodic dynamics for $\psi(i)$ is
\begin{equation}  \label{Eq1.36}
\frac{1}{M} \sum_{\mathbf{k}}
e^{\mathrm{i}\,\mathbf{k}(\mathbf{i}-\mathbf{j}
)}\,\Gamma(\mathbf{k})= \left\langle \psi(\mathbf{i})\right\rangle
\,\left\langle \psi^\dag(\mathbf{j})\right\rangle
\end{equation}
It is worth noting that $\Gamma(\mathbf{k})$ generally does not
assume its ergodic value [i.e., that required by
Eq.~(\ref{Eq1.36})] and has to be computed case by case according
to the dynamics and the boundary conditions. For instance, for any
finite system the statistical ensembles are not equivalent and the
criterion (\ref{Eq1.35}) holds only in the microcanonical one.
Moreover, the condition (\ref{Eq1.35}) is not satisfied by any
integral of motion or, more generally, by any operator that has a
diagonal part with respect to the Hamiltonian under
study\cite{Suzuki:71} (i.e., by any operator that has diagonal
entries whenever written in the basis of the eigenstates of the
Hamiltonian under study). This latter consideration clarifies why
the ergodic nature of the dynamics of an operator mainly depends
on the Hamiltonian it is subject to. It is really remarkable that
the \textbf{zero-frequency constants (\emph{ZFC})}, which are the
values of the zero-frequency function $\Gamma( \mathbf{k})$ over
the momenta for which $\mathcal{A}(\mathbf{k} )\neq\emptyset $,
are directly related to relevant measurable quantities such as the
compressibility, the specific heat, the magnetic
susceptibility,... For instance, we recall the formula that
relates the compressibility to the total particle number
fluctuations
\begin{equation}  \label{Eq1.36a}
\kappa=\beta \frac{M}{N^2} \left[\langle
\hat{N}^2\rangle-N^2\right]
\end{equation}
According to this, in the case of infinite systems too the correct
determination of the \emph{ZFC} cannot be considered as an
irrelevant issue (e.g., Eq.~(\ref{Eq1.36a}) holds in the
thermodynamic limit too). In conclusion, Eq.~(\ref{Eq1.36})
generally cannot be used to compute the \emph{ZFC}. In the next
section, we provide some examples of violation of the
condition~(\ref{Eq1.35}). It is necessary pointing out, in order
to avoid any possible confusion to the reader, that we are using
(here and in the examples presented in the next section)
\emph{full} operators and not \emph{fluctuation} ones (i.e., we
use operators not diminished of their average value, in contrast
with what it is usually done for the bosonic excitations like
spin, charge and pair). According to this, the \emph{ZFC} can be
different from zero (i.e., be equal to the squared average of the
operator), and still indicate an ergodic dynamics for the
operator.

Summarizing, by means of \emph{EM} and by using the boundary
conditions relative to the original definitions of the various
\emph{GF} we have been able to derive explicit expressions for
these latter [see Eqs.~(\ref{Eq1.18})]. However, these expressions
can only determine the functional dependence of the \emph{GF}:
their knowledge is not fully achieved yet. According to the
(anti)commutation relations, the normalization matrix $I^{(\eta
)}( \mathbf{k})$ usually contains some unknown functions that have
to be self-consistently calculated together with the \emph{ZFC}
(and the energy matrix $\varepsilon (\mathbf{k})$ if we use some
approximation scheme). These functions are static correlation
functions (correlators since now on) of operators not belonging to
the chosen basis. In principle, one could introduce a new set of
composite fields and repeat all scheme of calculations in order to
calculate the unknown correlators. However, the new set will
possibly generate other unknown correlators and the entire process
of self-consistency might become very cumbersome and, in most of
the cases, not convergent. An alternative scheme of calculation
can be proposed. Fixing the values of the unknown parameters and
of the \emph{ZFC} implies to put some constraints on the
representation where the \emph{GF} are realized. As the
determination of this representation is not arbitrary, it is clear
that there is no freedom in fixing these quantities. They must
assume values compatible with the dynamics and with the right
representation. Now, certainly the right representation must be
the one where all relations among the operators are systematically
conserved when the expectation values are taken (i.e., where all
the \emph{AC} and \emph{WT} are satisfied). It is then clear that
a shortcut in the procedure of self-consistency can be introduced.
We can fix the representation by requiring that
\begin{equation}
\left\langle \psi (i)\,\psi ^{\dag}(i)\right\rangle
=\frac{1}{M}\sum_{ \mathbf{k}}\frac{1}{2\pi}\int \!d\omega
\,C\left( \mathbf{k},\omega \right) \label{Eq1.37}
\end{equation}
where the l.h.s. is fixed by the \emph{AC}, the \emph{WT} and the
boundary conditions compatible with the phase under investigation
and in the r.h.s. the correlation function $C\left(
\mathbf{k},\omega \right) $ is computed by means of
Eq.~(\ref{Eq1.18c}). Equations~(\ref{Eq1.37}) generate a set of
self-consistent equations which determine the unknown parameters
(i.e., \emph{ZFC} and unknown correlators) and, consequently, the
proper representation\cite{Mancini-IJMPB,Mancini:98,Avella:98}. It
is worth noticing that the number of constraints generated by
Eqs.~(\ref{Eq1.37}) can be different from the number of unknowns
parameters. Generally, the coincidence of these two numbers
signals that the chosen basic set gives a reasonable description
of the dynamics contained in the truncated \emph{EM}.
Condition~(\ref{Eq1.37}) can be considered as a generalization, to
the case of composite fields, of the equation that, in the
non-interacting case, fixes the way of counting the particles per
site, according to the algebra, by determining the chemical
potential. According to this, the unknown correlators, coming from
the non-canonical (anti)commutation relations, have not be seen
like obstacles as many analytical techniques do, but like a
possibility to fix the representation and satisfy all the symmetry
relations. Any approximation not using them to do so will surely
fail in reproducing the physics of the system under study. It is
worth noting, and the examples of the next section will show how,
that by means of Eqs.~(\ref{Eq1.37}) is often possible to close
one sector (i.e., fermionic, spin, charge, pair, ...) at a time
without resorting to the opening of all or many of them
simultaneously. Obviously, this occurrence enormously facilitates
the calculations. Finally, it is worth noting that the entire
process of self-consistency [i.e., the use of Eqs.~(\ref{Eq1.37})]
will affect all the \emph{GF} at the same time and, therefore, all
the physical properties of the system. For instance, the linear
response of the system to an external perturbation
(susceptibility, conductivity, ...) is described by two-time
retarded \emph{GF}\cite{Kubo:57}. Although these type of \emph{GF}
do not explicitly depend on the \emph{ZFC}, there is an implicit
dependence through the internal self-consistent parameters, that
is the unknown correlators.

In this section, we have presented the general framework of the
\emph{COM}, which results to be a general method to deal with
composite fields and, consequently, with complex correlated
systems. In the next Section, we will illustrate this calculation
scheme by considering some specific examples.

\section{Examples}

\subsection{The two-site Hubbard model}

\label{Hub2}

The two-site Hubbard model is described by the following
Hamiltonian
\begin{equation}  \label{Eq3.1}
H=\sum_{ij}\left(t_{ij}-\delta_{ij}\,\mu\right)c^\dagger(i)\,c(j)
+ U\sum_i n_\uparrow(i)\,n_\downarrow(i)
\end{equation}
where the summation range only over two sites at distance $a$ from
each other and the rest of notation is standard\cite{Avella:01}.
The hopping matrix $t_{ij}$ is defined by
\begin{equation}  \label{Eq3.2}
t_{ij} = -2t\,\alpha_{ij} \;\; \;\; \;\; \alpha_{ij} =
\frac12\sum_k e^{ \mathrm{i}\,k(i-j)}\,\alpha(k)
\end{equation}
where $\alpha(k)=\cos(ka)$ and $k=0$, $\pi/a$.

We now proceed to study the system by means of the equation of
motion approach and the \emph{GF} formalism\cite{Avella:01}
described in Sec.~\ref {Forma}. A complete set of fermionic
eigenoperators of $\hat{H}$ is the following one
\begin{equation}  \label{Eq3.4}
\psi (i)=\left(
\begin{array}{l}
\xi (i) \\
\eta (i) \\
\xi _s(i) \\
\eta _s(i)
\end{array}
\right)
\end{equation}
where
\begin{subequations}
\begin{align}  \label{Eq3.5}
&\xi(i)=\left[1-n(i)\right]c(i) \\
&\eta(i)=n(i)\,c(i) \\
&\xi _s(i)=\frac12 \sigma ^\mu\,n_\mu (i)\,\xi ^\alpha (i)+\xi
(i)\,\eta
^{\dagger\alpha}(i)\,\eta (i) \\
&\eta _s(i)=\frac12 \sigma ^\mu\,n_\mu (i)\,\eta ^\alpha (i)+\xi
(i)\,\xi^{\dagger\alpha} (i)\,\eta (i)
\end{align}
We define $\psi^\alpha(i)=\sum_j \alpha_{ij}\,\psi(j)$ and use the
spinorial notation for the field operators. $n_\mu
(i)=c^\dagger(i)\,\sigma_\mu\,c(i)$ is the charge ($\mu=0$) and
spin ($\mu=1,2,3$) operator; greek (e.g., $\mu$, $\nu$) and latin
(e.g., $a$, $b$, $k$) indices take integer values from $0$ to $3$
and from $1$ to $3$, respectively; sum over repeated indices, if
not explicitly otherwise stated, is understood;
$\sigma_\mu=(1,\vec{\sigma})$ and $ \sigma^\mu=(-1,\vec{\sigma})$;
$\vec{\sigma}$ are the Pauli matrices. In momentum space the field
$\psi(i)$ satisfies the equation of motion
\end{subequations}
\begin{equation}  \label{Eq3.6}
\mathrm{i}\frac{\partial}{\partial t}\psi(k,t)=\varepsilon
(k)\,\psi (k,t)
\end{equation}
where the energy matrix $\varepsilon (k)$ has the expression
\begin{equation}  \label{Eq3.7}
\varepsilon(k)=\left(
\begin{array}{cccc}
-\mu -2t\,\alpha(k) & -2t\,\alpha(k) & -2t & -2t \\
0 & U-\mu & 2t & 2t \\
0 & 4t & -\mu +2t\,\alpha(k) & 4t\,\alpha(k) \\
0 & 2t & 2t\,\alpha(k) & U-\mu
\end{array}
\right)
\end{equation}

Straightforward calculations, according to the scheme traced in
Sec.~\ref {Forma}, show that two correlators
\begin{align}  \label{Eq3.8}
&\Delta = \left\langle \xi^\alpha(i)\,\xi^\dagger(i)\right\rangle
-
\left\langle \eta^\alpha(i)\,\eta^\dagger(i)\right\rangle \\
&p = \frac14 \left\langle n_\mu ^\alpha (i)\,n_\mu
(i)\right\rangle -\left\langle c_\uparrow (i)\,c_\downarrow
(i)\left[c_\downarrow ^\dagger (i)\,c_\uparrow ^\dagger
(i)\right]^\alpha\right\rangle
\end{align}
appear in the normalization matrix $I(\mathbf{k}) = \mathcal{F}
\left\langle \left\{\psi(\mathbf{i},t),\,
\psi^\dag(\mathbf{j},t)\right\}\right\rangle $. Then, the
\emph{GF} depend on three parameters: $\mu$, $\Delta$ and $p$. The
correlator $\Delta$ can be expressed in terms of the fermionic
correlation function $C(i,j)=\left\langle
\psi(i)\,\psi^\dagger(j)\right\rangle $; the chemical potential
$\mu$ can be related to the particle density by means of the
relation $n=2\left[1-C_{11}(i,i)-C_{22}(i,i)\right]$. The
parameter $p$ cannot be calculated in the fermionic sector; it is
expressed in terms of correlation functions of the bosonic fields
$n_\mu (i)$ and $c_\uparrow (i)\,c_\downarrow (i)$. According to
this, the determination of the fermionic \emph{GF} requires the
parallel study of bosonic \emph{GF}.

After quite cumbersome calculations, it is possible to
see\cite{Avella:01} that a complete set of bosonic eigenoperators
of $\hat{H}$ in the spin-charge channel is given by
\begin{equation}  \label{Eq3.9}
B^{(\mu )}(i)=\left(
\begin{array}{l}
B_1^{(\mu )}(i) \\
\vdots \\
B_6^{(\mu )}(i)
\end{array}
\right)
\end{equation}
where
\begin{align}  \label{Eq3.10}
&B_1^{(\mu )}(i)=c^\dagger (i)\,\sigma _\mu\,c(i) \\
&B_2^{(\mu )}(i)=c^\dagger (i)\,\sigma _\mu\,c^\alpha
(i)-c^{\dagger\alpha}
(i)\,\sigma _\mu\,c(i) \\
&B_3^{(\mu )}(i)=d_\mu (i)-d_\mu ^\alpha (i)+d_\mu ^\dagger
(i)-d_\mu
^{\dagger\alpha} (i) \\
&B_4^{(\mu )}(i)=d_\mu (i)-d_\mu ^\alpha (i)-d_\mu ^\dagger
(i)+d_\mu
^{\dagger\alpha} (i) \\
&B_5^{(\mu )}(i)=f_\mu (i)-f_\mu ^\alpha (i)-f_\mu ^\dagger
(i)+f_\mu
^{\dagger\alpha} (i) \\
&B_6^{(\mu)}(i)=f_\mu(i)-f_\mu^\alpha(i)+f_\mu^\dagger(i)-f_\mu^{\dagger
\alpha}(i)
\end{align}
with the definitions:
\begin{align}  \label{Eq3.11}
d_\mu (i)&=\xi ^\dagger (i)\,\sigma _\mu\,\eta ^\alpha (i) \\
f_0(i)&=-\eta ^\dagger (i)\,\eta (i)-d^\dagger (i)\,d^\alpha (i)  \notag \\
&+\eta ^\dagger (i)\,\eta (i)\,\xi ^{\dagger\alpha} (i)\,\xi ^\alpha (i) \\
f_a(i)&=\xi ^\dagger (i)\,\xi (i)\,n_a^\alpha (i)-\frac12
\mathrm{i} \,\epsilon_{abc}\,n_b(i)\,n_c^\alpha (i)
\end{align}
The field $B^{(\mu )}(i)$ satisfies the equation of motion
\begin{equation}  \label{Eq3.12}
\mathrm{i}\frac{\partial}{\partial t}B^{(\mu
)}(k,t)=\kappa(k)\,B^{(\mu )}(k,t)
\end{equation}
where the energy matrix $\kappa(k)$ has the expression
\begin{equation}  \label{Eq3.13}
\kappa (k)=\left(
\begin{array}{cccccc}
0 & -2t & 0 & 0 & 0 & 0 \\
-4t\left[1-\alpha(k)\right] & 0 & U & 0 & 0 & 0 \\
0 & 0 & 0 & U & 2t & 0 \\
0 & 0 & U & 0 & 0 & 2t \\
0 & 0 & 8t & 0 & 0 & 0 \\
0 & 0 & 0 & 8t & 0 & 0
\end{array}
\right)
\end{equation}

The energy spectra are given by
\begin{align}  \label{Eq3.14}
&\omega_1(k) = -2t \sqrt {2\left[1-\alpha (k)\right]} \\
&\omega_2(k) = 2t \sqrt {2\left[1-\alpha (k)\right]} \\
&\omega_3(k) = -U-4J_U \\
&\omega_4(k) = -4J_U \\
&\omega_5(k) = 4J_U \\
&\omega_6(k) = U+4J_U
\end{align}
where
\begin{equation}  \label{Eq3.15}
J_U = \frac18 \left[\sqrt {U^2+64t^2}-U \right]
\end{equation}

Straightforward calculations according to the scheme given in
Sec.~\ref {Forma} show that the correlation function has the
expression
\begin{multline}  \label{Eq3.16}
C^{(\mu )}(i,j) = \left\langle B^{(\mu )}(i)\,B^{(\mu
)\dagger}(j)\right\rangle \\
= \frac14 \sum_k \sum_{n=1}^6 e^{\mathrm{i}\,k(i-j)-\mathrm{i}
\,\omega_n(k)(t_i-t_j)} \\
\times \left[1+\tanh \frac{\beta\,\omega _n(k)}2 \right]
f^{(n,\mu)}(k)
\end{multline}
where
\begin{subequations}
\label{Eq3.17}
\begin{align}
&f^{(n,\mu )}(0) = 0 \;\; \;\; \mathnormal{for} \;\; n=3,4,5,6 \\
&f^{(n,\mu )}(\pi ) = \coth \frac{\beta\, \omega _n(\pi )}2 \sigma
^{(n,\mu )}(\pi )\;\;\forall n
\end{align}
\end{subequations}
Owing to the fact that zero-energy modes appear
for $n=1$, $2$ and $k=0$ [cfr. Eq.~(\ref{Eq3.14})], \emph{ZFC}
appear in the correlation functions
\begin{equation}  \label{Eq3.18}
\Gamma^{(\mu)}(0)=\frac12 \sum_{n=1}^2 f^{(n,\mu )}(0)
\end{equation}
In principle $\Gamma^{(\mu)}(0)$ could be calculated by means of
Eq.~(\ref {Eq1.16b}); however this would require the calculation
of the anticommutators $\left\langle \left\{B^{(\mu
)}(i,t),\,B^{(\mu )\dagger} (j,t)\right\}\right\rangle $ which
generate correlation functions of higher order giving raise to a
chain of \emph{GF} whose closure is not evident. Similar methods,
like the use of the relaxation function\cite{Kubo:57}, would lead
to the same problem. One might think, as is often done in the
literature, to fix this constant by its ergodic value. However,
this is not correct as we are in a finite system in the
grandcanonical ensemble and the ergodicity
condition~(\ref{Eq1.36}) does not hold. For the moment, we can
state that this constant remains undetermined.

The spectral density functions $\sigma ^{(n,\mu )}(k)$, calculated
by means of Eq.~(\ref{Eq1.9}) depends on a set of parameters which
come from the calculation of the normalization matrix $I^{(\mu
)}(k)=\mathcal{F} \left\langle \left[B^{(\mu )}(i,t),\,B^{(\mu
)\dagger} (j,t)\right] \right\rangle $. In particular, for the
(1,1)-component the following parameters appear:
\begin{subequations}
\label{Eq3.17b}
\begin{align}
&C_{12}^\alpha = \left\langle \eta ^\alpha (i)\,\xi ^\dagger
(i)\right\rangle
\\
&C^\alpha= \left\langle c^\alpha (i)\,c^\dagger (i)\right\rangle \\
&d= \left\langle c_\uparrow(i)\,c_\downarrow (i)\left[c_\downarrow
^\dagger
(i)\,c_\uparrow ^\dagger (i)\right]^\alpha \right\rangle \\
&\chi^\alpha _s = \left\langle \vec{n}(i)\cdot \vec{n}^\alpha
(i)\right\rangle
\end{align}
\end{subequations}
The parameters $C^\alpha$ and $C_{12}^\alpha$ are related to the
fermionic correlation function $C(i,j)=\left\langle
\psi(i)\,\psi^\dagger(j)\right\rangle $. The parameter
$\chi^\alpha _s$ can be expressed in terms of the bosonic
correlation function $ C^{(\mu)}(i,j)=\left\langle
B^{(\mu)}(i)\,B^{(\mu)\dagger}(j)\right\rangle$. In order to use
the standard procedure of self-consistency, we need to calculate
the parameter $d$. For this purpose we should open both the pair
channel and a double occupancy-charge channel (i.e., we will need
the static correlation function $\left\langle
n_\uparrow(i)\,n_\downarrow(i)\,n^\alpha(i)\right\rangle$). The
corresponding calculations are reported in
Ref.~\onlinecite{Avella:01} where is shown that these two channels
do not carry any new unknown \emph{ZFC}. The self-consistence
scheme closes; by considering the four channels (i.e., fermionic,
spin-charge, pair and double occupancy-charge) we can set up a
system of coupled self-consistent equations for all the
parameters. However, the \emph{ZFC} $\Gamma^{(\mu)}(0)$ has not
been determined yet: we have not definitely fixed the
representation of the \emph{GF}.

In conclusion, the standard procedure of self-consistency is very
involved and is not able to give a final answer because of the
problem of fixing the \emph{ZFC}. This problem is known in the
literature as the zero-frequency ambiguity of the \emph{GF}
formalism\cite{Callen:67,Zubarev:74,Huber:77,Aksenov:78,Aksenov:78a}.

We will now approach the problem by taking a different point of
view. The proper representation of the \emph{GF} must satisfy the
condition that all the microscopic laws, expressed as relations
among operators must hold also at macroscopic level as relations
among matrix elements. For instance, let us consider the fermionic
channel. We have seen that there exists the parameter $p$, not
explicitly related to the fermionic propagator, that can be
determined by opening other channels. However, we know that at the
end of the calculations, if the representation is the right one,
the parameter $p$ must take a value such that the
\emph{symmetries} are conserved. By imposing the \emph{AC}
(\ref{Eq1.37}) and by recalling the expression for $\Delta$ we get
three equations
\begin{subequations}
\begin{align}  \label{Eq3.25}
&n=2(1-C_{11}-C_{22}) \\
&\Delta =C_{11}^\alpha -C_{22}^\alpha \\
&C_{12}=0
\end{align}
\end{subequations}

This set of coupled self-consistent equations will allow us to
completely determine the fermionic \emph{GF}. Calculations
show\cite{Avella:01} that this way of fixing the representation is
the right one: all the \emph{symmetry} relations are satisfied and
all the results exactly agree with those obtained by means of
\emph{ED}. We do not have to open the bosonic channels; the
fermionic one is self-contained.

Next, let us consider the spin-charge \emph{GF}. In the
spin-charge sector we have the parameters $C^\alpha$,
$C_{12}^\alpha$, $\chi^\alpha _s$, $d$ and the two \emph{ZFC}
\begin{align}  \label{Eq3.26}
&b_0=\frac14 \sum_{i=1}^2 f_{11}^{(i,0)}(0) \\
&b_k=\frac14 \sum_{i=1}^2 f_{11}^{(i,k)}(0) \quad \quad k=1,2,3
\end{align}
Since we are in absence of an external applied magnetic field,
$b_k$ takes the same values for any value of $k$.

The parameter $C^\alpha$ and $C_{12}^\alpha$ are known, since the
fermionic correlation functions have been computed. The parameters
$\chi^\alpha _s$ and $d$ can be computed by means of the equations
\begin{align}  \label{Eq3.27}
&d=\frac14 \left\langle n_\mu ^\alpha (i)\,n_\mu (i)\right\rangle -p \\
&\chi^\alpha _s = \left\langle \vec{n}(i)\cdot \vec{n}^\alpha
(i)\right\rangle
\end{align}
The \emph{ZFC} are fixed by the \emph{AC}
\begin{equation}  \label{Eq3.28}
C_{11}^{(\mu )}(i,i)=\left\langle n_\mu (i)\,n_\mu
(i)\right\rangle
\end{equation}
By recalling (\ref{Eq3.16}) and (\ref{Eq3.17}) we have
\begin{equation}  \label{Eq3.29}
b_\mu =\left\langle n_\mu (i)\,n_\mu (i)\right\rangle
-\frac14\sum_{i=1}^6 \left[1+\coth \frac{\beta\,\omega _i(\pi
)}2\right]\sigma_{11}^{(i,\mu )}(\pi )
\end{equation}
with
\begin{equation}  \label{Eq3.30}
\left\langle n_\mu (i)\,n_\mu (i)\right\rangle =\left\{
\begin{array}{lll}
n+2D & \mathnormal{for} & \mu =0 \\
n-2D & \mathnormal{for} & \mu =1,2,3
\end{array}
\right.
\end{equation}
$D=\left\langle n_\uparrow (i)\,n_\downarrow (i)\right\rangle $ is
the double occupancy and can be calculated by means of the
fermionic correlation functions $D=n-1+C_{11}$.
Eqs.~(\ref{Eq3.27}) and (\ref{Eq3.29}) constitute a set of coupled
self-consistent equations which will determine completely the
Green's function in the spin-charge channel. Calculations show
that this way of fixing the representation is the right one: all
the symmetry relations are satisfied and all the results exactly
agree with those obtained by means of \emph{ED}.

\begin{figure}[t!]
\begin{center}
\includegraphics[width=8cm,keepaspectratio=true]{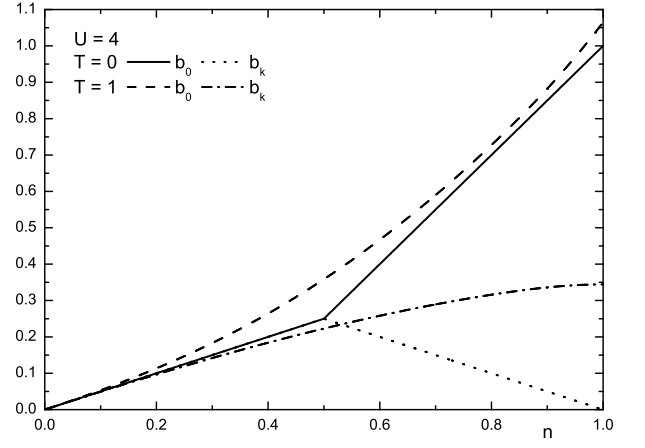}
\end{center}
\caption{$b_{0}$ and $b_{k}$ are plotted as functions of $n$ for
$U=4$ and $T=0$ and $1$. $U$ and $T$ are expressed in units of
$t$.} \label{Fig1}
\end{figure}

\begin{figure}[t!]
\begin{center}
\includegraphics[width=8cm,keepaspectratio=true]{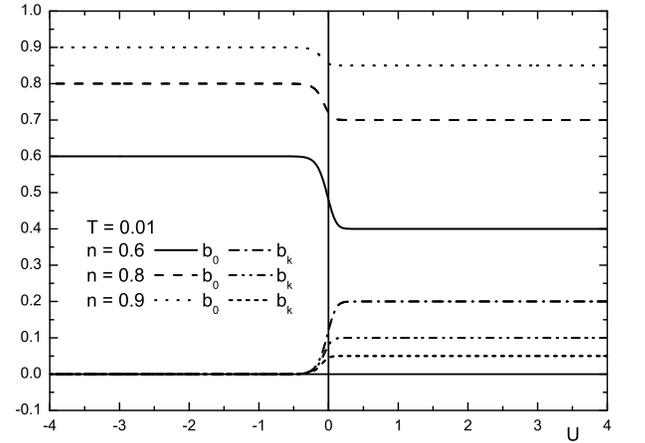}
\end{center}
\caption{$b_{0}$ and $b_{k}$ are plotted as functions of $U$ for
$T=0.01$ and $n=0.6$, $0.8$, and $0.9$. $U$ and $T$ are expressed
in units of $t$.} \label{Fig2}
\end{figure}

The \emph{ZFC} $b_0$ and $b_k$ are plotted as functions of $n$ and
$U$ in Figs.~\ref{Fig1} and \ref{Fig2}, respectively, for various
temperatures. It is worth noting that they assume their
\emph{ergodic} values (i.e. $n^2$ and $0$, respectively) only in
some regions of the parameter space: (at zero temperature) at
$n=1$ (both $b_{0}$ and $b_{k}$) and at $n=0.5$ ($b_{0}$ only). In
these regions, the grand-canonical ensemble is equivalent to the
microcanonical one and the underlying ergodicity of the charge and
spin dynamics emerges.

It is worth noting that the \emph{ZFC} $b_0$ is directly related
to the compressibility by means of the following
relation\cite{Avella:01}
\begin{equation}  \label{Eq3.31}
\kappa = \frac2{k_\mathrm{B}T}\frac1{n^2}\left[b_0-n^2\right]
\end{equation}
According to this, if we erroneously set the value of $b_0$ to the
ergodic one (i.e., $n^2$) we would get a constant zero
compressibility.

\subsection{The three-site Heisenberg model}

The three-site Heisenberg model, in presence of an external
magnetic field $ h $, is described by the following Hamiltonian
\begin{equation}
\hat{H}=J\sum\limits_{\mathbf{i}=1}^{3}\vec{S}(i)\cdot\vec{S}^{\alpha
}(i)-h\sum\limits_{\mathbf{i}=1}^{3}S_{z}(i)
\end{equation}
where $\vec{S}(i)$ is the local spin at the site $\mathbf{i}$,
with quantum number $S=\frac12$. The relative positions of the
three sites are those of the end-points and the middle-point of a
segment, in this case we are using periodic boundary conditions,
or those of the vertices of an equilateral triangle; in both
cases, the distances of two of them in the given order (i.e.,
$1\rightarrow2$, $2\rightarrow3$, $3\rightarrow1$) is taken to be
unitary. The notation $\vec{S}^{\alpha}(i)$ indicates
\begin{equation}
\vec{S}^{\alpha}(\mathbf{i},t)=\sum\limits_{\mathbf{j}}\alpha_{\mathbf{ij}
} \vec{S}(\mathbf{j},t)
\end{equation}
The projection operator $\alpha_{\mathbf{ij}}$ is defined by
\begin{align}
& \alpha_{\mathbf{ij}}={\frac{1}{3}}\sum\limits_{k}{}\mathrm{e}
^{\mathrm{i}
\mathbf{k(i-j)}}\alpha(\mathbf{k}) \\
& \alpha(\mathbf{k})=\cos(\mathbf{k})
\end{align}
where $\mathbf{k}=-{\frac{{2\pi}}{{3a}}}$, $0$,
${\frac{{2\pi}}{{3a}}}$.

A complete set of eigenoperators of $\hat{H}$ is
\begin{equation}
\psi^{(m)}(i)=\left(
\begin{array}{c}
\psi_{1}^{(m)}(i) \\
\psi_{2}^{(m)}(i) \\
\psi_{3}^{(m)}(i)
\end{array}
\right)
\end{equation}
where
\begin{align}
\psi_{1}^{(m)}(i) & =\left\{
\begin{tabular}{ll}
$S^{+}(i)=S_{x}+\mathrm{i}S_{y}$ & for$\;m=1$ \\
$S_{z}(i)$ & for$\;m=2$
\end{tabular}
\ \right. \\
\psi_{2}^{(m)}(i) & =\left\{
\begin{tabular}{ll}
$l^{+}(i)=l_{x}+\mathrm{i}l_{y}$ & for$\;m=1$ \\
$l_{z}(i)$ & for$\;m=2$
\end{tabular}
\ \right. \\
\psi_{3}^{(m)}(i) & =\left\{
\begin{tabular}{ll}
$u^{+}(i)=u_{x}+\mathrm{i}u_{y}$ & for$\;m=1$ \\
$u_{z}(i)$ & for$\;m=2$
\end{tabular}
\ \right.
\end{align}
Hereafter, we will use the two sets of indices $\{x,y,z\}$ and
$\{1,2,3\}$ interchangeably.

The composite fields $l_{k}(i)$ and $u_{k}(i)$ are defined as
\begin{align}
l_{k}(i) & =\mathrm{i}\epsilon_{kpq}S_{p}^{\alpha}(i)S_{q}(i) \\
u_{k}(i) & =\mathrm{i}\epsilon_{kpq}[l_{p}^{\alpha}(i)S_{q}
(i)+S_{p}^{\alpha}(i)l_{q}(i)]
\end{align}

The field $\psi^{(m)}(i)$ satisfies the equation of motion
\begin{equation}
\mathrm{i}\frac{\partial}{{\partial}t}\psi^{(m)}(i)=\varepsilon^{(m)}
\psi^{(m)}(i) \label{3sHEM}
\end{equation}
where the energy matrix $\varepsilon^{(m)}$ has the expression
\begin{equation}
\varepsilon^{(m)}=\left(
\begin{array}{ccc}
a_{m}h & 2J & 0 \\
0 & a_{m}h & 2J \\
0 & \frac{9}{8}J & a_{m}h
\end{array}
\right)
\end{equation}
with $a_{m}=1-\delta_{2m}$. The energy spectra are given by
\begin{align}
\omega_{1}^{(m)} & =a_{m}h \label{enezero}\\
\omega_{2}^{(m)} & ={\frac{1}{2}}(2a_{m}h-3J) \\
\omega_{3}^{(m)} & ={\frac{1}{2}}(2a_{m}h+3J)
\end{align}

By means of the equation of motion (\ref{3sHEM}), the correlation
function
\begin{multline}
\label{corr}
C^{(m)}(i,j)=\left\langle \psi^{(m)}(i)\psi^{\dagger(m)}(j)\right\rangle \\
={\frac{1}{3}}\sum\limits_{k}{\frac{1}{{2\pi}}}\int
d\omega\,\mathrm{e} ^{
\mathrm{i}\mathbf{k(i-j)}-\mathrm{i}\omega(t_{i}-t_{j})}C^{(m)}
(\mathbf{k} ,\omega)
\end{multline}
has the expression
\begin{equation}
C^{(m)}(\mathbf{k},\omega)=\sum\limits_{n=1}^{3}\delta\lbrack\omega-\omega
_{n}^{(m)}(\mathbf{k})]c^{(n,m)}(\mathbf{k})
\end{equation}
where the matrices $c^{(n,m)}(\mathbf{k})$ have to be calculated.

Straightforward calculations according to the scheme given in
Section II show that the correlation function is given by
\begin{multline}  \label{EqC1}
C^{(1)}(i,j)
={\frac{1}{6}}\sum\limits_{n=1}^{3}\sum\limits_{\mathbf{k}}
\mathrm{e}^{\mathrm{i}\mathbf{k(i-j)}-\mathrm{i}\omega_{n}^{(1)}
(k)(t_{i}-t_{j})} \\
\times[1+\coth\left( \frac{\beta\omega_{n}^{(1)}}{2}\right)
]\sigma^{(n,1)}( \mathbf{k})
\end{multline}
\begin{multline}  \label{EqC2}
C^{(2)}(i,j)
=b^{(2)}(\mathbf{i,j})+{\frac{1}{6}}\sum\limits_{n=2}^{3}
\sum\limits_{\mathbf{k}}\mathrm{e}^{\mathrm{i}\mathbf{k(i-j)}-\mathrm{i}
\omega_{n}^{(2)}(k)(t_{i}-t_{j})} \\
\times[1+\coth\left( \frac{\beta\omega_{n}^{(2)}}{2}\right) ]
\sigma^{(n,2)}( \mathbf{k})
\end{multline}
where the zero-frequency function
\begin{equation}
b^{(2)}(\mathbf{i,j})={\frac{1}{{2\pi}}}{\frac{1}{3}}\sum\limits_{\mathbf{k}
}\mathrm{e}^{\mathrm{i}\mathbf{k(i-j)}}c^{(1,2)}(\mathbf{k})
\end{equation}
appears owing to the presence of the zero-energy mode
${\omega_{1}^{(2)}=0}$.

The spectral density functions $\sigma^{(n,m)}(\mathbf{k})$,
calculated by means of Eq.~(\ref{Eq1.9}), have the following
expressions
\begin{align}
\sigma^{(1,m)}(\mathbf{k}) & =\lambda^{(1,m)}(\mathbf{k})A^{(1)} \label{magn1} \\
\sigma^{(2,m)}(\mathbf{k}) & =\lambda^{(2,m)}(\mathbf{k})A^{(2)} \\
\sigma^{(3,m)}(\mathbf{k}) & =\lambda^{(3,m)}(\mathbf{k})A^{(3)} \\
\lambda^{(1,m)}(\mathbf{k}) &
=I_{11}^{(m)}(\mathbf{k})-\frac{16}{9}
I_{22}^{(m)}(\mathbf{k}) \label{magn2} \\
\lambda^{(2,m)}(\mathbf{k}) &
=3I_{12}^{(m)}(\mathbf{k})-4I_{22}^{(m)}(
\mathbf{k}) \\
\lambda^{(3,m)}(\mathbf{k}) &
=3I_{12}^{(m)}(\mathbf{k})+4I_{22}^{(m)}( \mathbf{k})
\end{align}
with
\begin{align}
A^{(1)} & =\left(
\begin{array}{ccc}
1 & 0 & 0 \\
0 & 0 & 0 \\
0 & 0 & 0
\end{array}
\right) \\
A^{(2)} & =\left(
\begin{array}{ccc}
-\frac{2}{9} & \frac{1}{6} & -\frac{1}{8} \\
\frac{1}{6} & -\frac{1}{8} & \frac{3}{32} \\
-\frac{1}{8} & \frac{3}{32} & -\frac{9}{128}
\end{array}
\right) \\
A^{(3)} & =\left(
\begin{array}{ccc}
\frac{2}{9} & \frac{1}{6} & \frac{1}{8} \\
\frac{1}{6} & \frac{1}{8} & \frac{3}{32} \\
\frac{1}{8} & \frac{3}{32} & \frac{9}{128}
\end{array}
\right)
\end{align}

The normalization matrix
$I^{(m)}(\mathbf{k})=\mathcal{F}\left\langle [\psi
^{(m)}(\mathbf{i},t),\psi ^{\dagger
(m)}(\mathbf{j},t)]\right\rangle $, that has the form
\begin{align}
I^{(1)}(\mathbf{k})& =\left(
\begin{array}{ccc}
I_{11}^{(1)}(\mathbf{k}) & I_{12}^{(1)}(\mathbf{k}) &
I_{22}^{(1)}(\mathbf{k}
) \\
I_{12}^{(1)}(\mathbf{k}) & I_{22}^{(1)}(\mathbf{k}) &
\frac{9}{{16}}
I_{12}^{(1)}(\mathbf{k}) \\
I_{22}^{(1)}(\mathbf{k}) & \frac{9}{{16}}I_{12}^{(1)}(\mathbf{k})
& \frac{9}{ {16}}I_{22}^{(1)}(\mathbf{k})
\end{array}
\right) \\
I^{(2)}(k)& =\left(
\begin{array}{ccc}
0 & I_{12}^{(2)}(\mathbf{k}) & 0 \\
I_{12}^{(2)}(\mathbf{k}) & 0 & \frac{9}{{16}}I_{12}^{(2)}(\mathbf{k}) \\
0 & \frac{9}{{16}}I_{12}^{(2)}(\mathbf{k}) & 0
\end{array}
\right)
\end{align}
with
\begin{align}
I_{11}^{(1)}(\mathbf{k})& =2M \label{magn3} \\
I_{12}^{(1)}(\mathbf{k})& =-\left[ 1-\alpha (\mathbf{k})\right]
\left(
C_{11}^{(1)\alpha}+2C_{11}^{(2)\alpha}\right) \\
I_{12}^{(2)}(\mathbf{k})& =-\left[ 1-\alpha (\mathbf{k})\right]
C_{11}^{(1)\alpha} \\
I_{22}^{(1)}(\mathbf{k})& =-\left[ 1-\alpha (\mathbf{k})\right]
\left( {\frac{1}{4}}-{\frac{1}{4}}C_{11}^{(1)\alpha
}+{\frac{1}{2}}C_{11}^{(2)\alpha}-2C_{22}^{(1)}\right)
\end{align}
depends on the set of parameters
\begin{align}
& M=\left\langle S_{z}(i)\right\rangle =C_{11}^{(1)}-\frac{1}{2}
=\left\langle S^{+}(i)S^{-}(i)\right\rangle -\frac{1}{2} \\
& C_{11}^{(1)\alpha}=\left\langle S^{+\alpha}(i)S^{-}(i)\right\rangle \\
& C_{11}^{(2)\alpha}=\left\langle S_{z}^{\alpha}(i)S_{z}(i)\right\rangle \\
& C_{22}^{(1)}=-\left\langle l^{+}(i)l^{-}(i)\right\rangle
\end{align}

These parameters are expressed in terms of the correlation
function $ C^{(m)}(i,j)$ and self-consistent equations are easily
written by means of Eqs.~(\ref{EqC1}) and (\ref{EqC2}). However,
in order to close the set of equations we need to know the
zero-frequency constant
\begin{equation}
b_{11}^{(2)\alpha}={\frac{1}{3}}\sum\limits_{\mathbf{k}}\alpha
(\mathbf{k} )b_{11}^{(2)}(\mathbf{k})={\frac{1}{{2\pi
}}}{\frac{1}{3}}\sum\limits_{ \mathbf{k}}\alpha
(\mathbf{k})c_{11}^{(1,2)}(\mathbf{k})
\end{equation}

This quantity, undetermined within the bosonic sector, can be
obtained, as proposed in Section II, by fixing the representation
of the GF by means of Eq.~(\ref{Eq1.37}). In particular, the
\emph{AC} requires that
\begin{equation}
\left\langle S^{+}(i)l^{-}(i)\right\rangle
={\frac{1}{2}}M+\left\langle S_{z}^{\alpha
}(i)S_{z}(i)\right\rangle +{\frac{1}{2}}\left\langle S^{+\alpha
}(i)S^{-}(i)\right\rangle
\end{equation}

This equation, together with the others coming from the
definitions (\ref {EqC1}) and (\ref{EqC2}), gives a set of five
coupled self-consistent equation for the five parameters $M$,
$C_{11}^{(1)\alpha}$, $ C_{11}^{(2)\alpha}$, $C_{22}^{(1)}$,
$b_{11}^{(2)\alpha}$. The system can be analytically solved. In
particular, the magnetization per site $M$ and the zero-frequency
constant $b_{11}^{(2)}$ are given by
\begin{align}
M& =\frac{1}{6}\tanh \left( \frac{\beta h}{2}\right)
\frac{{2+3\cosh (}\beta h{)+\mathrm{e}^{\beta
\frac{3}{2}J}}}{{\mathrm{e}^{\beta \frac{3}{2}J}+\cosh
(}\beta h{)}} \\
b_{11}^{(2)\alpha}& =\frac{9\cosh (\beta h)-4-\mathrm{e}^{\beta
\frac{3}{2} J}}{36\left[ \mathrm{e}^{\beta \frac{3}{2}J}+\cosh
(\beta h)\right]}
\end{align}

It should be noted that other \emph{ZFC} appear into the model.
Again, they can be fixed by the Algebra Constraint. For example,
the \emph{ZFC}
\begin{equation}
b_{11}^{(2)}={\frac{1}{3}}\sum\limits_{\mathbf{k}}b_{11}^{(2)}(\mathbf{k})={
\ \frac{1}{{2\pi
}}}{\frac{1}{3}}\sum\limits_{\mathbf{k}}c_{11}^{(1,2)}(
\mathbf{k})
\end{equation}
is determined by means of the equation
\begin{equation}
\left\langle S_{z}(i)S_{z}(i)\right\rangle ={\frac{1}{4}}
\end{equation}
and takes the value
\begin{equation}
b_{11}^{(2)}=\frac{9\cosh (\beta h)-4+5\mathrm{e}^{\beta
\frac{3}{2}J}}{36 \left[ \mathrm{e}^{\beta \frac{3}{2}J}+\cosh
(\beta h)\right]}
\end{equation}

\begin{figure}[t]
\begin{center}
\includegraphics[width=8cm,keepaspectratio=true]{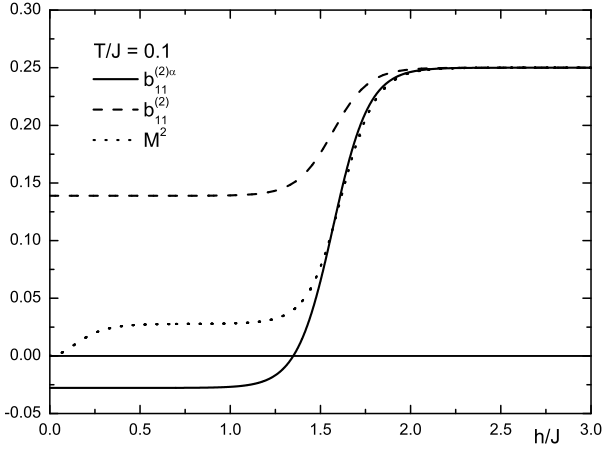}
\end{center}
\caption{The zero-frequency constants $b_{11}^{(2)\protect\alpha
}$ and $ b_{11}^{(2)}$ are plotted as a function of the magnetic
field for $T/J=0.1$. For comparison, the ergodic value $M^{2}$ is
also given.} \label{Fig3}
\end{figure}

\begin{figure}[t]
\begin{center}
\includegraphics[width=8cm,keepaspectratio=true]{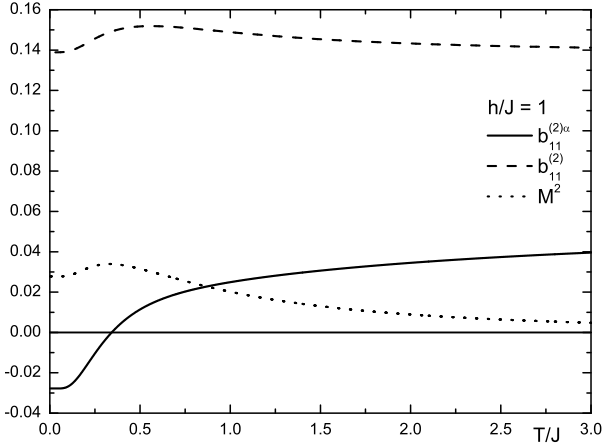}
\end{center}
\caption{The zero-frequency constant $b_{11}^{(2)\protect\alpha}$
and $ b_{11}^{(2)}$ are plotted as a function of the temperature
for $h/J=1$. For comparison, the ergodic value $M^{2}$ is also
given.} \label{Fig4}
\end{figure}

We note that the two \emph{ZFC} $b_{11}^{(2)\alpha}$ and
$b_{11}^{(2)}$ assume the ergodic value $M^{2}$ only in the limit
of very large external magnetic field $h$. In Figs.~\ref{Fig3} and
\ref{Fig4} the zero-frequency constants $b_{11}^{(2)\alpha}$ and
$b_{11}^{(2)}$ are plotted as a function of the magnetic field
$h/J$ and temperature $T/J$, respectively. For comparison the
ergodic value $M^{2}$ is also reported.

Let us now consider the $S^{\pm}$-channel (i.e., $m=1$) and the
relative \emph{ZFC}. For $h\neq \frac{3}{2}J$ there are no
zero-energy modes and the \emph{ZFC} relative to the $S^{\pm
}$-operators assume the ergodic value, i.e., zero. For the special
case $h=\frac{3}{2}J$ a zero-energy mode $\omega _{2}^{(1)}=0$
appears and the \emph{ZFC} become nonergodic. Again, the Algebra
Constraint can be used to fix these quantities. Straightforward
calculations, according to the proposed scheme, give
\begin{multline}
b_{11}^{(1)}={\frac{1}{3}}\sum\limits_{\mathbf{k}}b_{11}^{(1)}(\mathbf{k})={
\frac{1}{{2\pi
}}}{\frac{1}{3}}\sum\limits_{\mathbf{k}}c_{11}^{(2,1)}(
\mathbf{k})  \notag \\
=\frac{3\mathrm{e}^{\beta \frac{3}{2}J}+4}{18\mathrm{e}^{-3\beta
J}\left( 1+ \mathrm{e}^{\beta \frac{3}{2}J}\right) ^{3}}
\end{multline}

\begin{multline}
b_{11}^{(1)\alpha}={\frac{1}{3}}\sum\limits_{\mathbf{k}}\alpha
(\mathbf{k} )b_{11}^{(1)}(\mathbf{k})={\frac{1}{{2\pi
}}}{\frac{1}{3}}\sum\limits_{
\mathbf{k}}\alpha (\mathbf{k})c_{11}^{(2,1)}(\mathbf{k})  \notag \\
=-\frac{3\mathrm{e}^{\beta \frac{3}{2}J}+1}{9\mathrm{e}^{-3\beta
J}\left( 1+ \mathrm{e}^{\beta \frac{3}{2}J}\right) ^{3}}
\end{multline}

Let us now consider the limit of zero temperature. From the
previous expressions it is easy to derive the following.

\begin{itemize}
\item Case $J>0$ (Antiferromagnetic exchange)
\begin{equation}
M=
\begin{cases}
\frac{1}{2} & \text{for }h>\frac{3}{2}J \\
\frac{1}{6} & \text{for }h<\frac{3}{2}J
\end{cases}
\end{equation}
\end{itemize}

The two \emph{ZFC} $b_{11}^{(2)\alpha}$ and $b_{11}^{(2)}$ take
the values
\begin{eqnarray}
b_{11}^{(2)} &=&b_{11}^{(2)\alpha}=\frac{1}{4}\qquad \text{for
}h>\frac{3}{
2}J \\
b_{11}^{(2)} &=&-5b_{11}^{(2)\alpha}=\frac{5}{36}\qquad \text{for
}h<\frac{ 3}{2}J
\end{eqnarray}

They are ergodic for $h>\frac{3}{2}J$ and nonergodic for
$h<\frac{3}{2}J$ . The other two \emph{ZFC} $b_{11}^{(1)\alpha}$
and $b_{11}^{(1)}$ take the ergodic value ($0$) for any $h>0$. In
the limit of $h\rightarrow 0$ we obtain the ferromagnetic
solution.

\begin{itemize}
\item Case $J<0$ (Ferromagnetic exchange)
\begin{equation}
M=\frac{1}{2}
\end{equation}
\end{itemize}

All the zero-frequency constants take the ergodic value. In the
limit of $h\rightarrow 0$ we obtain again the ferromagnetic
solution.

Let us now consider the case of absence of external magnetic field
($h=0$). For this situation there are two energy modes, one in the
$S_{z}$-channel (i.e., $m=2$) $\omega _{1}^{(2)}=0$ and one in the
$S^{\pm}$-channel (i.e., $m=1$) $\omega _{1}^{(1)}=0$. In order to
avoid divergencies in the correlation functions, it must be
$\sigma ^{(1,m)}(\mathbf{k})=0$ for $m=1,$ $2$ and for all values
of $\mathbf{k}$. It must be
\begin{eqnarray}
I_{11}^{(m)}(\mathbf{k}) &=&0\Rightarrow M=0 \\
I_{22}^{(m)}(\mathbf{k}) &=&0\Rightarrow {1}-C_{11}^{(1)\alpha
}+{2} C_{11}^{(2)\alpha}-8C_{22}^{(1)}=0
\end{eqnarray}

By solving the self-consistent equations and by means of the
Algebra Constraint, one finds the following nonergodic values for
the \emph{ZFC}:

\begin{eqnarray}
b_{11}^{(2)} &=&\frac{1}{2}b_{11}^{(1)}=\frac{5}{36} \\
b_{11}^{(2)\alpha} &=&\frac{1}{2}b_{11}^{(1)\alpha
}=\frac{5-\mathrm{e} ^{\beta \frac{3}{2}J}}{36\left(
\mathrm{e}^{\beta \frac{3}{2}J}+1\right)}
\end{eqnarray}

At zero temperature
\begin{eqnarray}
b_{11}^{(2)} &=&\frac{1}{2}b_{11}^{(1)}=\frac{5}{36} \\
b_{11}^{(2)\alpha} &=&\frac{1}{2}b_{11}^{(1)\alpha}=
\begin{cases}
-\frac{1}{36} & \text{for}J>0 \\
\frac{5}{36} & \text{for}J<0
\end{cases}
\end{eqnarray}

Summarizing, we have
\begin{equation*}
\begin{tabular}{||l|l|l|l|l|l||}
\hline\hline $T$ & $h$ & $J$ & $M$ & $S_{z}$ & $S^{\pm}$ \\
\hline\hline $0$ & $\rightarrow 0$ & $>0$ & $\frac{1}{6}$ & N & E
\\ \hline $0$ & $\rightarrow 0$ & $<0$ & $\frac{1}{2}$ & E & E \\
\hline $\rightarrow 0$ & $0$ & $\neq 0$ & $0$ & N & N \\ \hline
$0$ & $\neq 0$ & $<0$ & $\frac{1}{2}$ & E & E \\ \hline $0$ &
$<\frac{3}{2}J$ & $>0$ & $\frac{1}{6}$ & N & E \\ \hline $0$ &
$>\frac{3}{2}J$ & $>0$ & $\frac{1}{2}$ & E & E \\ \hline $\neq 0$
& $0$ & $\neq 0$ & $0$ & N & N \\ \hline $\neq 0$ & $\rightarrow
\infty $ & $\neq 0$ & $\frac{1}{2}$ & E & E \\ \hline $\neq 0$ &
$\neq 0,\frac{3}{2}J$ & $\neq 0$ & $\neq 0$ & N & E \\ \hline
$\neq 0$ & $\frac{3}{2}J$ & $\neq 0$ & $\neq 0$ & N & N \\
\hline\hline
\end{tabular}
\end{equation*}
where E and N stays for an ergodic and nonergodic behavior of the
corresponding \emph{ZFC}, respectively. The first two lines of the
table consider the cases in which we have a ferromagnetic
solution.

The operator $S_{z}=\sum_{\mathbf{i}}S_{z}(i)$, as any constant of
motion, has essentially a non ergodic dynamics. Actually, for high
values of $h/J$, $S_{z}$ is forced to assume the higher possible
value $\frac{3}{2}$ and no fluctuations are allowed (i.e., the
susceptibility vanishes): the dynamics returns to be ergodic. At
zero temperature and for the ferromagnetic case (i.e., $J<0$) the
system is polarized and $S_{z}$ is ergodic for all finite value of
the magnetic field and in the ferromagnetic phase. The operator
$S^{+}=\sum_{\mathbf{i}}S^{+}(i)$ has an ergodic dynamics only in
presence of the magnetic field $h$ or in the ferromagnetic phase
as it is no longer an integral of motion in these cases. Also, for
the special case $h=\frac{3}{2}J$ the operator $S^{+}$ becomes
nonergodic. These results show how the ergodicity of the dynamics
of an operator can strongly depend on the boundary conditions.

It is worth noticing that the ferromagnetic phase has been
obtained only at exactly zero temperature (i.e., when the applied
magnetic field has been sent to zero after setting the temperature
to zero). This is due to the size of the system; finite systems
can sustain ordered phases only at exactly zero temperature. The
correlation functions in direct space should be computed by finite
sums over momenta (see Eq.~\ref{corr}) and for vanishing spectra
(e.g., for vanishing applied magnetic field; see
Eq.~\ref{enezero}) the Bose factor (see the $\coth$ in
Eq.~\ref{EqC1}) diverges except at exactly zero temperature. Only
in this latter case (i.e., $T=0$) the corresponding spectral
density function can retain a finite value (and consequently the
magnetization too; see Eqs.~\ref{magn1}, \ref{magn2} and
\ref{magn3}) instead of being forced to vanish in order to avoid
divergences in the direct space correlation functions. In
practice, we allow the magnetization to be finite and search for a
fully self-consistent solution. The system will self-adjust by
selecting only those states with a finite magnetization of the
same \emph{sign} of that assigned as initial condition according
to the ergodicity breaking inherent to any symmetry breaking.

We wish to remark that all the results obtained in this section
exactly agree with those obtained by means of \emph{ED}.

\subsection{A narrow-band Bloch system in presence of an external magnetic
field}

A narrow-band Bloch system in presence of an external magnetic
field is described by the following Hamiltonian
\begin{equation}  \label{Eq4.1}
H=\sum_{\bf ij}\left(t_{\bf ij}-\mu\,\delta_{\bf
ij}\right)c^\dagger (i)\,c(j)-h\sum_{\bf i} n_3(i)
\end{equation}
where $n_3(i)$ is the third component of the spin density operator
and $h$ is the intensity of the external magnetic field. The
indices $\bf i$ and $\bf j$ run on an infinite $d$-dimensional
lattice. Straightforward calculations show that the causal Green's
function $G^{(\mu )}_C(i,j)=\left\langle
\mathcal{T}\left[n_\mu(i)\,n_\mu(j)\right] \right\rangle $ and the
correlation function $C^{(\mu )}(i,j)=\left\langle
n_\mu(i)\,n_\mu(j) \right\rangle$ of the charge-spin operator $
n_\mu(i)=c^\dagger(i)\,\sigma_\mu\,c(i)$ have the following
expressions
\begin{align}
G^{(\mu )}_C(\mathbf{k},\omega )&=-\mathrm{i}\,(2\pi)^{d+1}
a^{-d}\,\delta^{(d)}(k)\, \delta (\omega)\, \Gamma^{(\mu)}
-Q^{(\mu )}(
\mathbf{k},\omega )  \label{Eq4.2} \\
C^{(\mu )}(\mathbf{k},\omega )&=(2\pi )^{d+1}\,a^{-d}\,\delta
^{(d)}(k)\,
\delta (\omega )\,\Gamma^{(\mu)}  \notag \\
&+\left[1+\tanh \frac{\beta\, \omega}2\right] \Im \left[Q^{(\mu
)}(\mathbf{ \ k},\omega )\right] \label{Eq4.3}
\end{align}
where $\delta^{(d)}(k)$ is the $d$-dimensional Dirac delta
function. $Q^{(\mu )}(\mathbf{k},\omega )$ comes from the proper
fermionic loop and is the Fourier transform of
\begin{equation}  \label{Eq4.4}
Q^{(\mu )}(i,j)=\Tr\left[\sigma _\mu\, G_C(i,j)\,\sigma _\mu\,
G_C(j,i) \right]
\end{equation}

Here $G_C(i,j)=\left\langle
\mathcal{T}\left[c(i)\,c^\dagger(j)\right] \right\rangle$ is the
causal fermionic function and has the expression
\begin{align}  \label{Eq4.5}
G_C(\mathbf{k},\omega )&=\sum_{n=1}^2 \frac{\sigma
^{(n)}}{1+e^{-\beta\,E_n(
\mathbf{k})}}  \notag \\
& \times \left[\frac1{\omega -E_n(\mathbf{k})+\mathrm{i}\delta} +
\frac{ e^{-\beta\,E_n(\mathbf{k})}} {\omega
-E_n(\mathbf{k})-\mathrm{i}\delta} \right]
\end{align}
with
\begin{align}
&E_1(\mathbf{k})=-\mu -2d\,t\,\alpha (\mathbf{k})-h \\
&E_2(\mathbf{k})=-\mu -2d\,t\,\alpha (\mathbf{k})+h  \label{Eq4.6} \\
&\sigma ^{(1)}=\left(
\begin{array}{cc}
1 & 0 \\
0 & 0
\end{array}
\right) \;\; \;\; \sigma ^{(2)}=\left(
\begin{array}{cc}
0 & 0 \\
0 & 1
\end{array}
\right) \label{Eq4.7}
\end{align}
where
\begin{equation}  \label{Eq6.4}
\alpha (\mathbf{k})=\frac1d \sum_{i=1}^d \cos (k_i\,a)
\end{equation}

The \emph{ZFC} $\Gamma^\mu$ is fixed by the \emph{AC}
(\ref{Eq1.37}) which requires
\begin{align}  \label{Eq4.8}
&\Gamma^{(\mu)} = \left\langle n_\mu (i)\,n_\mu (i)\right\rangle  \notag \\
&- \frac{a^d}{(2\pi )^{d+1}} \int \! d^dk \, d\omega \left[1+\tanh
\frac{ \beta \, \omega}2\right]\Im \left[Q^{(\mu
)}(\mathbf{k},\omega )\right]
\end{align}

The loop $Q^{(\mu )}(\mathbf{k},\omega )$ can be calculated by
means of (\ref {Eq4.5}). Calculations show
\begin{align}  \label{Eq4.9}
& \frac{a^d}{(2\pi )^{d+1}} \int \! d^dk \, d\omega \left[1+\tanh
\frac{
\beta \, \omega}2\right]\Im [Q^{(\mu )}(\mathbf{k},\omega )]  \notag \\
& =\left\langle n\right\rangle -\left\langle
n_\uparrow\right\rangle ^2-\left\langle n_\downarrow\right\rangle
^2 \;\; \;\; \mathnormal{for} \;\;
\mu=0,3 \\
& =\left\langle n\right\rangle -2\left\langle n_\uparrow
(i)\right\rangle \left\langle n_\downarrow (i)\right\rangle \;\;
\;\; \mathnormal{for} \;\; \mu=1,2
\end{align}

By recalling the \emph{AC} (\ref{Eq3.30}), Eq.~(\ref{Eq4.8}) gives
for the \emph{ZFC}
\begin{align}  \label{Eq4.10}
& \Gamma^{(0)} = \left\langle n\right\rangle ^2 \\
& \Gamma^{(1,2)} = 0 \\
& \Gamma^{(3)} = \left\langle n_3\right\rangle ^2
\end{align}
in accordance with the ergodic nature of the spin and charge
dynamics in this system.

It is worth noting that the compressibility of this system can be
computed by means of the general formula (\ref{Eq1.36a}) that
holds in the thermodynamic limit too and gives
\begin{equation}
\kappa =\frac{1}{\left\langle n\right\rangle ^{2}}\frac{\beta
}{2}\frac{ a^{d}}{2(2\pi )^{d}}\sum_{n=1}^{2}\int
\!d^{d}k\frac{1}{C_{n}(\mathbf{k})} \label{Eq4.10a}
\end{equation}
where $C_{n}(\mathbf{k})=\cosh ^{2}\left( \frac{\beta
\,E_{n}(\mathbf{k})}{2} \right) $. We can see that an ergodic
charge dynamics can lead to a non-ergodic value of the \emph{ZFC}
relative to the total number operator, which is an integral of
motion. Also in the infinite systems the decoupling inspired by
the requirement of ergodicity cannot always be applied.

\subsection{The Double Exchange Model}

The Double Exchange Model is defined by the following
Hamiltonian\cite {Zener:51}
\begin{equation}  \label{Eq6.1}
H=\sum_{\bf ij}\left(t_{\bf ij}-\mu \, \delta_{\bf
ij}\right)c^\dagger(i) \, c(j) - J_ \mathrm{H}\sum_{\bf i}
\vec{s}(i) \cdot \vec{S}(i)
\end{equation}
$\vec{s}(i)$ is the spin density operator of the electron and is
given by $ \vec{s}(i)= \frac12 c^\dagger (i) \, \vec{\sigma} \,
c(i)$; $\vec{S}(i)$ is a localized spin; $J_\mathrm{H}$ is the
ferromagnetic Hund coupling ($J_ \mathrm{H}>0$). In the
nearest-neighbor approximation for a $d$-dimensional cubic lattice
with lattice constant $a$, $t_{\bf ij}$ takes the form
\begin{equation}  \label{Eq6.3}
t_{\bf ij}=-2d\,t\,\alpha_{\bf ij}=-2d\,t\frac1N \sum_\mathbf{k}
e^{\mathrm{i}\, \mathbf{k}\cdot (\mathbf{i}-\mathbf{j})}\,\alpha
(\mathbf{k})
\end{equation}
where $\alpha (\mathbf{k})$ has been defined in the previous
section and $d$ is the dimensionality of the system. Let us
introduce the Heisenberg field
\begin{equation}  \label{Eq6.5}
B(i)=\left(
\begin{array}{l}
s^+(i) \\
S^+(i)
\end{array}
\right)
\end{equation}
where $s^\pm(i)$ and $S^\pm(i)$ are the standard rising and
lowering spin operators.

This field satisfy the equation of motion
\begin{equation}  \label{Eq6.6}
J(i) =\mathrm{i}\frac\partial{\partial t} B(i) = \left(
\begin{array}{l}
2d\,t\,\rho(i) + J_\mathrm{H}\,\lambda (i) \\
-J_\mathrm{H}\,\lambda (i)
\end{array}
\right)
\end{equation}
where
\begin{align}  \label{Eq6.7}
&\rho(i) = c_\uparrow ^{\dagger\alpha}(i) \, c_\downarrow(i) -
c_\uparrow^\dagger(i) \, c_\downarrow^\alpha(i) \\
&\lambda(i) = s^+(i)\,S_z(i)-s_z(i)\,S^+(i)
\end{align}
We linearize the equation of motion (\ref{Eq6.6}) by projecting
the source $ J(i)$ on the basis (\ref{Eq6.5})
\begin{equation}  \label{Eq6.8}
J(i) \approx \sum_\mathbf{j} \varepsilon^B (\mathbf{i},\mathbf{j})
\, B( \mathbf{j},t)
\end{equation}
where the coefficients are determined by the following equation
\begin{equation}  \label{Eq6.9}
\left\langle \left[J(\mathbf{i},t),\,B^\dagger
(\mathbf{j},t)\right] \right\rangle = \sum_\mathbf{l}
\varepsilon^B (\mathbf{i},\mathbf{l}) \left\langle
\left[B(\mathbf{l},t),\,B^\dagger (\mathbf{j},t)\right]
\right\rangle
\end{equation}

Let us compute, within the framework described in
Sec.~\ref{Forma}, the causal Green's function
\begin{align}
& G(\mathbf{k},\omega )=\mathcal{F}\left\langle \mathcal{T}\left[
B(i)\,B^{\dagger}(j)\right] \right\rangle   \notag  \label{Eq6.10} \\
& =\sum_{n=1}^{2}\left[ P\left( \frac{\sigma
^{(n)}(\mathbf{k})}{\omega -\omega _{n}(\mathbf{k})}\right)
-\mathrm{i}\,\pi \,\delta \left[ \omega -\omega
_{n}(\mathbf{k})\right] g^{(n)}(\mathbf{k})\right]
\end{align}
and the correlation function
\begin{equation}
C(\mathbf{k},\omega )=\mathcal{F}\left\langle B(i)\,B^{\dagger
}(j)\right\rangle =\sum_{n=1}^{2}\delta \left[ \omega -\omega
_{n}(k)\right] c^{(n)}(\mathbf{k})
\end{equation}
$g^{(n)}(\mathbf{k})$ and $c^{(n)}(\mathbf{k})$ are still unknown
functions; $\omega _{n}(\mathbf{k})$ are the eigenvalues of the
matrix $\varepsilon ^{B}(\mathbf{k})$; $\sigma ^{(n)}(\mathbf{k})$
are the density spectral functions, completely determined by the
matrices $\varepsilon ^{B}(\mathbf{k} )$ and
$I^{B}(k)=\mathcal{F}\left\langle \left[ B(i,t),\,B^{\dagger
}(j,t) \right] \right\rangle $ by means of relation (\ref{Eq1.9}).
We have
\begin{equation}
I^{B}(k)=\left(
\begin{array}{cc}
2\left\langle s_{z}(i)\right\rangle  & 0 \\
0 & 2\left\langle S_{z}(i)\right\rangle
\end{array}
\right)
\end{equation}

For the sake of brevity, the explicit expressions for the energy
matrix $ \varepsilon ^{B}(\mathbf{k})$, the energy spectra $\omega
_{n}(\mathbf{k})$ and the spectral density functions $\sigma
^{(n)}(\mathbf{k})$ are not reported here and can be found in
Ref.~\onlinecite{Mancini:00e}. The calculations show that
\begin{eqnarray}
\lim_{\mathbf{k}\rightarrow \mathbf{0}}\omega _{2}(\mathbf{k}) &=&0 \\
\lim_{\mathbf{k}\rightarrow \mathbf{0}}\sigma ^{(2)}(\mathbf{k})
&=&\frac{ I_{11}^{B}I_{22}^{B}}{I_{11}^{B}+I_{22}^{B}}\left(
\begin{array}{cc}
1 & -1 \\
-1 & 1
\end{array}
\right)
\end{eqnarray}

According to the scheme of calculation given in Sec.~\ref{Forma},
we generally have
\begin{align}
& c^{(n)}(\mathbf{k})=2\pi \,\delta _{n2}\,\delta
(\mathbf{k})\,\Gamma ^{B}+\pi \left[ 1+\coth \frac{\beta \omega
_{n}(\mathbf{k})}{2}\right]
\sigma ^{(n)}(\mathbf{k})  \label{Eq6.13} \\
& g^{(n)}(\mathbf{k})=2\delta _{n2}\,\delta (\mathbf{k})\,\Gamma
^{B}+\coth \frac{\beta \omega _{n}(\mathbf{k})}{2}\sigma
^{(n)}(\mathbf{k})
\end{align}

The Green's functions have the following expressions
\begin{multline}
G(\mathbf{k},\omega )=-2\,\mathrm{i}\,\delta (\omega )\,\delta
(\mathbf{k}
)\,\Gamma ^{B}  \notag \\
+\sum\limits_{n=1}^{2}\frac{\sigma ^{(n)}(\mathbf{k})}{1-e^{-\beta
\omega}} \left[ \frac{1}{\omega -\omega
_{n}(\mathbf{k})+\mathrm{i}\,\delta}-\frac{ e^{-\beta \,\omega
}}{\omega -\omega _{n}(\mathbf{k})-\mathrm{i}\,\delta} \right]
\end{multline}

\begin{multline}
C(\mathbf{k},\omega )=2\pi \,\delta (\omega )\,\delta
(\mathbf{k})\,\Gamma
^{B}  \notag \\
+\pi \sum\limits_{n=1}^{2}\delta \left[ \omega -\omega
_{n}(\mathbf{k}) \right] \left[ 1+\coth \frac{\beta \,\omega
_{n}(\mathbf{k})}{2}\right] \sigma ^{(n)}(\mathbf{k})
\end{multline}

The \emph{ZFC} $\Gamma ^{B}$ is determined by means of the
\emph{AC} which requires
\begin{multline}
\frac{a^{d}}{(2\pi )^{d}}\Gamma ^{B}=\left\langle B(i)\,B^{\dagger
}(i)\right\rangle  \label{Eq6.14} \\
-\frac{a^{d}}{2(2\pi )^{d}}\sum_{n=1}^{2}\int \!d^{d}k\left[
1+\coth \frac{ \beta \,\omega _{n}(\mathbf{k})}{2}\right] \sigma
^{(n)}(\mathbf{k})
\end{multline}

In the case of a three-dimensional system the integral in
Eq.~(\ref{Eq6.14}) is finite. The Green's functions are fully
determined and a ferromagnetic order does exist. In the case $d<3$
we must distinguish two cases.

\begin{itemize}
\item[$T>0$] In this case the divergence of the integrand in
Eq.~(\ref {Eq6.14}) is not integrable and the integral is
divergent. The only physical solution is absence of ferromagnetic
order. The magnetization must vanish
\begin{eqnarray}
&&I_{11}^{B}=2\left\langle s_{z}(i)\right\rangle =0 \\
&&I_{22}^{B}=2\left\langle S_{z}(i)\right\rangle =0
\end{eqnarray}
The spectral density function $\sigma ^{(2)}(\mathbf{k})$
vanishes, in agreement with the general relation (\ref{Eq1.16c}).

\item[$T=0$] In this case Eq.~(\ref{Eq6.14}) becomes
\begin{multline}
\frac{a^{d}}{(2\pi )^{d}}\Gamma ^{B}=\left\langle B(i)\,B^{\dagger
}(i)\right\rangle\\
-\frac{a^{d}}{(2\pi )^{d}}\sum_{n=1}^{2}\int \!d^{d}k\,\theta
\!\left[ \omega _{n}(\mathbf{k})\right] \sigma ^{(n)}( \mathbf{k})
\end{multline}
\end{itemize}
where $\theta \!\left[ \cdots \right]$ is the ordinary step
function. We see that we can have ferromagnetic order for any
dimension.

The results of the calculations for this model illustrate what
stated in Section II regarding broken symmetry states in bulk
systems. When a zero-energy mode appears in the ladder operator
sector (e.g., $S^+$) the corresponding spectral density function,
which is related to the order parameter (e.g., $\sigma \propto I
\propto \langle S^z \rangle$), must vanish for all finite systems
at finite temperatures and for all infinite systems when the
divergence of the Fourier coefficients of the correlation function
is not integrable in order to avoid divergences in the direct
space correlation functions. At zero temperature it is always
possible to have a finite value for the spectral density function
and consequently for the order parameter. This result is a
manifestation of the Mermin-Wagner theorem\cite{Mermin:66} which
prevents the system from approaching, in some dimension, a
particular ordered phase except at zero temperature.

\section{Conclusions}

In conclusion, the \emph{GF} formalism for composite operators has
been revised by making use of the equations of motion method. It
has been shown that all the general relations (spectral
representation, sum rules, etc.) can be derived without resorting
to the knowledge of the complete set of eigenstates of the
Hamiltonian. The advantage of using the equations of motion
formalism is that it can be applied to any operatorial basis both
exact and approximate. Special attention has been paid to the
presence of the \emph{ZFC} and to the problem of determining
unknown parameters related to higher order correlators. The
\emph{ZFC} issue is quite relevant because such quantities are
directly related to many response functions. We have shown that an
effective and proper way to fix the representation is to impose
the constraints coming from the \emph{AC} and the \emph{WT}. When
these conditions are required, a set of self-consistent equations
is obtained that permits to compute both the parameters appearing
in the spectral functions and the zero-frequency component of the
\emph{GF}, avoiding the problem of \emph{uncontrolled} and
\emph{uncontrollable} decouplings, which affects many different
approximation schemes and has been here definitely solved.

Moreover, it is worth reminding the following issues, which have
been discussed in detail all over the text:
\begin{itemize}
 \item The two-time retarded (advanced) and causal bosonic \emph{GF} carry substantially different information.
 \item The ergodicity condition cannot be used a priori to compute the \emph{ZFC}.
 \item The Mermin-Wagner theorem\cite{Mermin:66} naturally appears as a requirement to avoid divergences in the direct space correlation functions.
\end{itemize}

It is also necessary pointing out -- we already did it in
Section~II -- that, although a careful choice of the components
for the basic set makes possible the description of the main
scales of energy present in the system under analysis, the
inclusion of fully momentum and frequency dependent self-energy
corrections can be sometime necessary to take into account
low-energy and virtual processes.

The calculation scheme has been illustrated by considering four
systems: the two-site Hubbard model, the three site Heisenberg
system, the narrow-band Bloch system and the Double-Exchange
model. These examples clearly show the relevance and complexness
of the above issues and illustrate in detail the application of
the proposed procedure. It has been checked that the proposed
scheme gives the exact result when solvable systems are
considered.

\begin{acknowledgments}
We would like to thank Prof. N.M. Plakida for many enlightening
discussions and for helping us in placing the ergodicity issue in
its proper historical context. We also wish to thank Prof. H.
Matsumoto, Prof. A.E. Ruckenstein and Prof. V. Srinivasan for
their very useful comments and remarks on the manuscript and the
stimulating discussions.
\end{acknowledgments}

\appendix

\section{Generalized perturbative approach for \emph{SCES}}

Given a certain Hamiltonian
$\hat{H}=\hat{H}\left[\varphi(i)\right]$, where $\varphi(i)$
denotes an Heisenberg electronic field
[$i=\left(\mathbf{i},t\right)$] in spinorial notation satisfying
canonical anticommutation relations, and a set of composite
operators $\psi(i)$ chosen in the spirit of the discussion given
in Section I, the equations of motion for the propagator of the
field $\psi(i)$ can be obtained by the dynamics obeyed by this
latter which reads as
\begin{equation}
\mathrm{i}\frac\partial{\partial
t}\psi(i)=\left[\psi(i),H\right]=J(i)
\end{equation}

In complete generality, this equation can be rewritten as
\begin{equation}\label{EMexp}
\mathrm{i}\frac\partial{\partial t}\psi(i) = \sum_\mathbf{j}
\varepsilon(\mathbf{i},\mathbf{j}) \psi(\mathbf{j},t) +
\delta\!J(i)
\end{equation}
where the linear term $\varepsilon\,\psi$ represents the
projection of the source $J(i)$ on the basis $\psi(i)$. The energy
matrix $\varepsilon(\mathbf{i},\mathbf{j})$ can be computed by
means of the equation
\begin{equation}
\left\langle \left[ \delta\!J(\mathbf{i},t),
\psi^\dagger(\mathbf{j},t) \right]_\eta \right\rangle = 0
\end{equation}
which defines the residual source $\delta\!J(i)$ and gives
\begin{equation}
\varepsilon(\mathbf{i},\mathbf{j})=\sum_\mathbf{l} \left\langle
\left[ J(\mathbf{i},t), \psi^\dagger(\mathbf{l},t) \right]_\eta
\right\rangle \left\langle \left[\psi(\mathbf{l},t),
\psi^\dagger(\mathbf{j},t) \right]_\eta \right\rangle^{-1}
\end{equation}

Obviously, also less systematic projections of the source could be
attempted and will result in different determinations of
$\varepsilon(\mathbf{i},\mathbf{j})$ and $\delta\!J(i)$.

After Eq.~(\ref{EMexp}), the Fourier transform $G_Q^{(\eta
)}(\mathbf{k},\omega )$ of the \emph{GF} $G_Q^{(\eta )}(i,j)$,
where $Q=R$ (retarded), $A$ (advanced), $C$ (causal) (see
definitions in Section II), satisfies the following equation
\begin{multline}
G_Q^{(\eta )}(\mathbf{k},\omega )=G_{Q,0}^{(\eta
)}(\mathbf{k},\omega )\\
+G_{Q,0}^{(\eta )}(\mathbf{k},\omega )\left[I^{(\eta
)}(\mathbf{k})\right]^{-1}\Sigma_Q^{(\eta )}(\mathbf{k},\omega
)G_Q^{(\eta )}(\mathbf{k},\omega ) \label{Dyson}
\end{multline}
where the propagator $G_{Q,0}^{(\eta )}(\mathbf{k},\omega )$ is
defined by the equation
\begin{equation}
\left[ \omega - \varepsilon (\mathbf{k}) \right] G_{Q,0}^{(\eta
)}(\mathbf{k},\omega )=I^{(\eta )}(\mathbf{k})
\end{equation}

The matrix $I^{(\eta )}(\mathbf{k})$, known as the normalization
matrix, is defined as
\begin{equation}
I^{(\eta )}(\mathbf{k}) = \mathcal{F} \left\langle \left[
\psi(\mathbf{i},t), \psi^\dagger(\mathbf{j},t) \right]_\eta
\right\rangle
\end{equation}
$\Sigma_Q^{(\eta )}(\mathbf{k},\omega )$ is the proper self-energy
and has the expression
\begin{equation}
\Sigma_Q^{(\eta )}(\mathbf{k},\omega ) =
B_{Q,irr}^{(\eta)}(\mathbf{k},\omega )I^{(\eta)}(\mathbf{k})^{-1}
\end{equation}
where $B_{Q,irr}^{(\eta)}(\mathbf{k},\omega )$ is the irreducible
part of the propagator $B_Q^{(\eta)}(\mathbf{k},\omega) =
\mathcal{F} \left\langle \mathcal{Q} \left[ \delta\!J(i)
\delta\!J^\dagger(j)\right]\right\rangle$. Equation~(\ref{Dyson})
can be formally solved to give
\begin{equation}\label{solution}
G_Q^{(\eta )}(\mathbf{k},\omega
)=\frac{1}{\omega-\varepsilon(\mathbf{k})-\Sigma_Q^{(\eta
)}(\mathbf{k},\omega )}I^{(\eta)}(\mathbf{k})
\end{equation}
Equations~(\ref{Dyson}) and (\ref{solution}) are nothing else than
the Dyson equation for a formulation based on composite fields and
represents the starting point for a perturbative calculation in
terms of the propagator $G_{Q,0}^{(\eta )}(\mathbf{k},\omega )$.
The properties and the determination of this latter are derived
and discussed in Section II.

\section{\emph{KMS} relation and the general formulation}

From the definitions~(\ref{Eq1.2})-(\ref{Eq1.4}) we can derive the
following exact relations
\begin{subequations}
\label{Eq1.11}
\begin{multline}
G^{(\eta)}_R(i,j)+G^{(\eta)}_A(i,j) = 2G^{(\eta )}_C(i,j) \\
-\left\langle \left[\psi(i),\,\psi^\dag(j)\right]_{-\eta
}\right\rangle
\end{multline}
\begin{equation}
G^{(\eta)}_R(i,j)-G^{(\eta)}_A(i,j) = \left\langle
\left[\psi(i),\,\psi^ \dag(j)\right]_\eta\right\rangle
\end{equation}
\end{subequations}

By making use of the Kubo-Martin-Schwinger (\emph{KMS}) relation $
\left\langle A(t)\,B(t^{\prime})\right\rangle =\left\langle
B(t^{\prime})\,A(t+\mathrm{i}\,\beta)\right\rangle $, where $A(t)$
and $B(t)$ are Heisenberg operators at time $t$, the
$\eta$-commutator can be expressed in terms of the correlation
function as
\begin{multline}  \label{Eq1.12}
\left\langle \left[\psi(i),\,\psi^\dag(j)\right]_\eta\right\rangle
= \frac{1}{M} \sum_\mathbf{k} \frac{1}{2\pi} \int\!d\omega \,
e^{\mathrm{i}\,\mathbf{ \ k
}\cdot\left(\mathbf{i}-\mathbf{j}\right)-\mathrm{i}\,\omega
\left(t_i-t_j\right)} \\
\times\left[1+\eta\,e^{-\beta\,\omega}\right]C\left(\mathbf{k},\omega\right)
\end{multline}
where $M$ is the number of sites and $\mathbf{k}$ runs over the
first Brillouin zone. Then, the equations~(\ref{Eq1.11}) in
momentum space become
\begin{subequations}
\label{Eq1.13}
\begin{multline}
\sum_{l=1}^n \delta\!\left[\omega -\omega_l(\mathbf{k})\right]
\left\{
g^{(\eta,l)}_C(\mathbf{k}) \right. \\
\left. -\frac{1}{2\pi}
\left[1-\eta\,e^{-\beta\,\omega}\right]\,c^{(l)}(
\mathbf{k})\right\}=0
\end{multline}
\begin{multline}
\sum_{l=1}^n \delta\!\left[\omega -\omega_l(\mathbf{k})\right]
\left\{
\sigma^{(\eta,l)}(\mathbf{k}) \right. \\
\left.-\frac{1}{2\pi}
\left[1+\eta\,e^{-\beta\,\omega}\right]\,c^{(l)}(
\mathbf{k})\right\}=0
\end{multline}
\end{subequations}

In order to solve these equations, we have to take into account
that for any given momentum $\mathbf{k}$ we can always write
\begin{equation}\label{momreg}
\omega _{l}(\mathbf{k})=
\begin{cases}
=0 & \text{for $l\in \mathcal{A}(\mathbf{k})\subseteq \aleph
=\{1,\ldots ,n\}
$} \\
\neq 0 & \text{for $l\in \mathcal{B}(\mathbf{k})=\aleph
-\mathcal{A}(\mathbf{ k})$}
\end{cases}
\end{equation}
Obviously, $\mathcal{A}(\mathbf{k})$ can also be the empty set
(i.e., $ \mathcal{A}(\mathbf{k})=\emptyset $ and
$\mathcal{B}(\mathbf{k})=\aleph $). Combined use of
Eqs.~(\ref{Eq1.13}) and (\ref{momreg}) gives the results reported
in Sec.~II [Eq.~(\ref{Eq1.16})].

\section{Formulation in the limit of zero temperature}

At zero temperature Eq.~(\ref{Eq1.13}) is not applicable and we
should proceed in the following way (the usual derivation in terms
of a complete set of eigenstates of the Hamiltonian can be found
in Ref.~[\onlinecite{Tyablikov:67}]).

Let us consider the correlation functions
\begin{multline}
C_{\psi \psi ^{\dagger}}(i,j)=\left\langle \psi (i)\psi ^{\dagger
}(j)\right\rangle   \notag \\
=\frac{1}{N}\sum\limits_{k}\frac{1}{2\pi}\int d\omega
\,e^{\mathrm{i}\left[ \mathbf{k}\cdot
(\mathbf{i}-\mathbf{j})-\omega (t_{i}-t_{j})\right]}C_{\psi \psi
^{\dagger}}(\mathbf{k},\omega )
\end{multline}
\begin{multline}
C_{\psi ^{\dagger}\psi}(i,j)=\left\langle \psi ^{\dagger}(i)\psi
(j)\right\rangle   \notag \\
=\frac{1}{N}\sum\limits_{k}\frac{1}{2\pi}\int d\omega
\,e^{\mathrm{i}\left[ \mathbf{k}\cdot
(\mathbf{i}-\mathbf{j})-\omega (t_{i}-t_{j})\right]}C_{\psi
^{\dagger}\psi}(\mathbf{k},\omega )
\end{multline}

By taking the limit $T\longrightarrow 0$ of the \emph{KMS}
relation
\begin{equation}
C_{\psi ^{\dagger}\psi}(\mathbf{k},\omega )=e^{-\beta \omega
}C_{\psi \psi ^{\dagger}}(\mathbf{k},\omega )
\end{equation}
it is immediate to see that for any finite value of the Fourier
coefficients, it must be
\begin{eqnarray}
C_{\psi \psi ^{\dagger}}(\mathbf{k},\omega ) &=&
\begin{cases}
\neq 0 & \text{for }\omega \geq 0 \\
=0 & \text{for }\omega <0
\end{cases}
\\
C_{\psi ^{\dagger}\psi}(\mathbf{k},\omega ) &=&
\begin{cases}
=0 & \text{for }\omega >0 \\
\neq 0 & \text{for }\omega \leq 0
\end{cases}
\end{eqnarray}

Furthermore
\begin{equation}
C_{\psi ^{\dagger}\psi}(\mathbf{k},0)=C_{\psi \psi ^{\dagger
}}(\mathbf{k} ,0)
\end{equation}

Let us consider the energy spectra $\omega _{l}(\mathbf{k})$ and
let us write in complete generality, for any given momentum
$\mathbf{k}$
\begin{equation}
\omega _{l}(\mathbf{k})=\left\{
\begin{array}{ll}
=0 & \text{for $l\in \mathcal{A}(\mathbf{k})\subseteq \aleph $} \\
>0 & \text{for $l\in \mathcal{C}(\mathbf{k})\subseteq \aleph $} \\
<0 & \text{for $l\in \mathcal{D}(\mathbf{k})\subseteq \aleph $}
\end{array}
\right.
\end{equation}

Then, Eqs.~(\ref{Eq1.11}) in momentum space are written as
\begin{subequations}
\begin{multline}
\delta \!(\omega )\sum_{l\in \mathcal{A}(\mathbf{k})}\left\{
g_{C}^{(\eta ,l)}(\mathbf{k})-\frac{1}{2\pi}\left( 1-\eta \right)
c_{\psi \psi ^{\dagger
}}^{(l)}(\mathbf{k})\right\} \\
+\sum_{l\in \mathcal{C}(\mathbf{k})}\delta \!\left[ \omega -\omega
_{l}( \mathbf{k})\right] \left\{ g_{C}^{(\eta
,l)}(\mathbf{k})-\frac{1}{2\pi}
c_{\psi \psi ^{\dagger}}^{(l)}(\mathbf{k})\right\} \\
+\sum_{l\in \mathcal{D}(\mathbf{k})}\delta \!\left[ \omega -\omega
_{l}( \mathbf{k})\right] \left\{ g_{C}^{(\eta
,l)}(\mathbf{k})+\frac{\eta}{2\pi} c_{\psi ^{\dagger}\psi
}^{(l)}(\mathbf{k})\right\} =0
\end{multline}
\begin{multline}
\sum_{l=1}^{n}\delta \!\left[ \omega -\omega
_{l}(\mathbf{k})\right] \left\{ \sigma ^{(\eta
,l)}(\mathbf{k})-\frac{1}{2\pi}\left( 1+\eta \right) c_{\psi
\psi ^{\dagger}}^{(l)}(\mathbf{k})\right\} \\
+\sum_{l=1}^{n}\delta \!\left[ \omega -\omega
_{l}(\mathbf{k})\right] \left\{ \sigma ^{(\eta
,l)}(\mathbf{k})-\frac{1}{2\pi}c_{\psi \psi
^{\dagger}}^{(l)}(\mathbf{k})\right\} \\
+\sum_{l=1}^{n}\delta \!\left[ \omega -\omega
_{l}(\mathbf{k})\right] \left\{ \sigma ^{(\eta
,l)}(\mathbf{k})-\frac{\eta}{2\pi}c_{\psi ^{\dagger}\psi
}^{(l)}(\mathbf{k})\right\} =0
\end{multline}
\end{subequations}

The solution of these equations gives the following expressions
for the \emph{GF}
\begin{subequations}
\label{solT0}
\begin{equation}
G_{R,A}^{(\eta )}(\mathbf{k},\omega )=\sum_{l\in \aleph
}\frac{\sigma ^{(\eta ,l)}(\mathbf{k})}{\omega -\omega
_{l}(\mathbf{k})\pm \mathrm{i} \,\delta}
\end{equation}
\begin{multline}
G_{C}^{(\eta )}(\mathbf{k},\omega )=\Gamma (\mathbf{k})\left(
\frac{1}{ \omega +\mathrm{i}\,\delta}+\frac{\eta}{\omega
-\mathrm{i}\,\delta}
\right) \\
+\sum_{l\notin \mathcal{A}(\mathbf{k})}\sigma ^{(\eta
,l)}(\mathbf{k})\left[ \frac{\theta \!\left[ \omega
_{l}(\mathbf{k})\right]}{\omega -\omega _{l}(
\mathbf{k})+\mathrm{i}\,\delta}+\frac{\theta \!\left[ -\omega
_{l}(\mathbf{ k})\right]}{\omega -\omega
_{l}(\mathbf{k})-\mathrm{i}\,\delta}\right]
\end{multline}
\begin{multline}
C_{\psi \psi ^{\dagger}}(\mathbf{k},\omega )=2\pi \,\delta (\omega
)\,\Gamma (\mathbf{k}) \\
+2\pi \sum_{l\notin \mathcal{A}(\mathbf{k})}\delta \!\left[ \omega
-\omega _{l}(\mathbf{k})\right] \theta \!\left[ \omega
_{l}(\mathbf{k})\right] \sigma ^{(\eta ,l)}(\mathbf{k})
\end{multline}
\begin{multline}
C_{\psi ^{\dagger}\psi}(\mathbf{k},\omega )=2\pi \,\delta (\omega
)\,\Gamma (\mathbf{k}) \\
+2\pi \eta \sum_{l\notin \mathcal{A}(\mathbf{k})}\delta \!\left[
\omega -\omega _{l}(\mathbf{k})\right] \theta \!\left[ -\omega
_{l}(\mathbf{k}) \right] \sigma ^{(\eta ,l)}(\mathbf{k})
\end{multline}
\end{subequations}
It is fairly easy to check that these expressions (\ref{solT0})
correspond to limit $T \rightarrow 0$ ($\beta \rightarrow \infty$)
of the expressions (\ref{Eq1.18}).

\section{Useful relations}

We note the dispersion relations

\begin{subequations}
\begin{multline}  \label{Eq1.19}
\Re\left[G^{(\eta)}_{R,A}(\mathbf{k},\omega )\right]= \\
\mp \frac{1}{\pi} \mathcal{P}\left\{\int \!
d\omega^{\prime}\frac{1} {\omega -\omega^{\prime}}
\Im\left[G^{(\eta )}_{R,A}(\mathbf{k},\omega^{\prime})
\right]\right\}
\end{multline}
\begin{multline}  \label{Eq1.20}
\Re\left[G^{(\eta)}_C(\mathbf{k},\omega )\right]= \\
- \frac{1}{\pi} \mathcal{P}\left\{\int \! d\omega^{\prime}\frac{1}
{\omega -\omega^{\prime}} \frac{1+\eta\,
e^{-\beta\,\omega^{\prime}}}{1-\eta\,
e^{-\beta\,\omega^{\prime}}}\Im\left[G^{(\eta )}_C(\mathbf{k}
,\omega^{\prime})\right]\right\}
\end{multline}
\end{subequations}
This latter relation is valid for causal \emph{fermionic}
\emph{GF} [i.e., for $\eta=1$] only when $\Gamma (\mathbf{k})=0$.

For the retarded and advanced \emph{GF}, which are analytical
functions satisfying the standard Kramers-Kronig relations
(\ref{Eq1.19}), we can establish a spectral representation
\begin{equation}  \label{Eq1.21}
G^{(\eta )}_{R,A}(\mathbf{k},\omega )=\int \!
d\omega^{\prime}\frac{\rho ^{(\eta)
}(\mathbf{k},\omega^{\prime})}{\omega -\omega^{\prime}\pm
\mathrm{i} \,\delta}
\end{equation}
where we introduced the spectral function
\begin{equation}  \label{Eq1.22}
\begin{split}
\rho ^{(\eta )}(\mathbf{k},\omega
)&=\sum_{l=1}^n\delta\!\left[\omega
-\omega _l(\mathbf{k})\right]\sigma ^{(\eta,l)}(\mathbf{k}) \\
&=\mp \frac{1}{\pi}\Im\!\left[G^{(\eta )}_{R,A}(\mathbf{k},\omega
)\right]
\end{split}
\end{equation}
A spectral representation for the causal \emph{GF} can be
established in the following form
\begin{equation}
G^{(\eta )}_{C}(\mathbf{k},\omega )=\int \! d\omega^{\prime}\frac{
\rho ^{(\eta) }(\mathbf{k},\omega^{\prime})}{1+\eta \,e^{-\beta
\,\omega ^{\prime }}}\left( \frac{1}{\omega -\omega ^{\prime
}+\mathrm{i}\,\delta }+ \frac{\eta\,e^{-\beta \,\omega ^{\prime
}}}{\omega -\omega ^{\prime }-\mathrm{i}\,\delta }\right)
\end{equation}
This latter relation is valid for causal \emph{bosonic} \emph{GF}
[i.e., for $\eta=-1$] only when $\Gamma (\mathbf{k})=0$.

We also note the sum rule
\begin{equation}  \label{Eq1.23}
\int \! d\omega \, \rho ^{(\eta )}(\mathbf{k},\omega
)=\sum_{l=1}^n \sigma ^{(\eta,l)}(\mathbf{k})=I^{(\eta
)}(\mathbf{k})
\end{equation}
This is a particular case of the general sum rule
\begin{equation}  \label{Eq1.24}
\begin{split}
\int \! d\omega \, \omega^p \, \rho ^{(\eta )}(\mathbf{k},\omega
)&=\sum_{l=1}^n \omega_l^p(\mathbf{k}) \, \sigma ^{(\eta,l)}(\mathbf{k}) \\
&=M^{(\eta,p)}(\mathbf{k})=\varepsilon^p\,I
\end{split}
\end{equation}
where $M^{(\eta,p)}(\mathbf{k})$ are the spectral moments
\begin{equation}  \label{Eq1.25}
M^{(\eta,p)}(\mathbf{k})=\mathcal{F}\left[
\mathrm{i}\frac{\partial^p}{
\partial \, t^p_i} \left\langle \left[\psi(i) , \, \psi^\dag(j)\right]_\eta
\right\rangle \right]_{t_i=t_j}
\end{equation}
and the last equality in Eq.~(\ref{Eq1.24}) holds only when
Eq.~(\ref{Eq2.7} ) also does.

Finally, by exploiting the independence of $c^{(l)}(\mathbf{k})$
on $\eta $ [see Eq.(\ref{Eq1.16d})], we have
\begin{equation}
\sigma ^{(-1,l)}(\mathbf{k})=\tanh {\frac{\beta \,\omega
_{l}(\mathbf{k})}{2}}\sigma
^{(1,l)}(\mathbf{k})\,\,\,\,\,\,\forall l\in
\mathcal{B}(\mathbf{k}) \label{Eq1.16bis}
\end{equation}
In absence of symmetry breaking, Eqs.~(\ref{Eq1.16c}) and
(\ref{Eq1.16bis}), together with Eq.~(\ref{Eq1.9}), give
\begin{equation}
I_{\alpha \beta}^{(-1)}(\mathbf{k})=\sum_{l\delta}\Omega _{\alpha
l}( \mathbf{k})\tanh {\frac{\beta \,\omega
_{l}(\mathbf{k})}{2}}\Omega _{l\delta
}^{-1}(\mathbf{k})\,I_{\delta \beta}^{(1)}(\mathbf{k})
\label{Eq1.17ter}
\end{equation}
The independence of $C(\mathbf{k},\omega )$ on $\eta $ [see
Eq.(\ref{Eq1.34} )] gives
\begin{equation}
\Im \left[ G_{C}^{(\eta )}(\mathbf{k},\omega )\right] =\Im \left[
G_{R}^{(-\eta )}(\mathbf{k},\omega )\right]
\end{equation}
In terms of spectral densities
\begin{equation}
\sum_{l\in \mathcal{A}(\mathbf{k})}\delta \!\left[ \omega -\omega
_{l}( \mathbf{k})\right] \left[ \frac{\sigma
^{(1,l)}(\mathbf{k})}{1+e^{-\beta \,\omega
_{l}(\mathbf{k})}}-\frac{\sigma ^{(-1,l)}(\mathbf{k})}{1-e^{-\beta
\,\omega _{l}(\mathbf{k})}}\right] =0
\end{equation}

\bibliographystyle{apsrev}
\bibliography{biblio}

\begin{thebibliography}{40}
\bibitem[{Pau()}]{Pauli1}
\bibinfo{note}{It is also possible to have fermionic systems with occupation
  numbers per site equal to either 0 or 1 independently from the spin (e.g.,
  the $t$-$J$ model) or just fixed to 1 (e.g., the spin-$\frac12$ Heisenberg
  model)}.
\bibitem[{rec()}]{recipe}
\bibinfo{note}{We can choose: the higher order fields emerging from the
  equations of motion (i.e., the conservation of some spectral moments is
  assured)\cite{Mancini:98b}, the eigenoperators of some relevant interacting
  terms (i.e., the relevant interactions are correctly
  treated)\cite{Avella:02e}, the eigenoperators of the problem reduced to a
  small cluster\cite{Avella:03b}, the composite field describing the Kondo-like
  singlet emerging at low-energy in any electronic spin
  system\cite{Villani:00},...}
\bibitem[{bas()}]{basis}
\bibinfo{note}{We choose fermionic operators to study the local, thermodynamic
  and single-particle properties and bosonic ones to analyze the response
  functions.}
\end{thebibliography}

\end{document}